\documentclass[aps, prd, nofootinbib, superscriptaddress]{revtex4-2}
\pdfoutput=1
\usepackage{amsmath, amssymb, graphicx, hyperref, braket}
\usepackage{caption, subcaption}
\usepackage[english]{babel}
\usepackage[dvipsnames]{xcolor}
\usepackage{soul}
\newcommand{\mbx}{\mathbf{x}}

\newcommand{\be}[1]{ \begin{equation} }
	\newcommand{\ee}{\end{equation}}
\newcommand{\bea}[1]{\begin{eqnarray} }
	\newcommand{\eea}{\end{eqnarray}}
\newcommand{\bes}{\begin{subequations}}
	\newcommand{\ees}{\end{subequations}}

\newcommand{\p}{\partial}

\newcommand{\w}{\omega}

\newcommand{\non}{\nonumber}

\def\deltabar{{\mathchar '26\mkern -10mu\delta}}
\begin{document}
	\title{One-Loop quantum effects in Carroll scalars}
	\author{Kinjal Banerjee}
	\email{kinjalb@gmail.com}
	\affiliation{Department of Physics, BITS-Pilani, K K Birla Goa Campus, Zuarinagar, Goa 403726, India.}
	\author{Rudranil Basu}
	\email{rudranilb@goa.bits-pilani.ac.in}
	\affiliation{Department of Physics, BITS-Pilani, K K Birla Goa Campus, Zuarinagar, Goa 403726, India.}
	\author{Bhagya Krishnan}
	\email{p20190007@goa.bits-pilani.ac.in}
	\affiliation{Department of Physics, BITS-Pilani, K K Birla Goa Campus, Zuarinagar, Goa 403726, India.}
	\author{Sabyasachi Maulik}
	\email{mauliks@iitk.ac.in}
	\affiliation{Saha Institute of Nuclear Physics, A CI of Homi Bhabha National Institute, 1/AF, Bidhannagar, West Bengal 700064, India.}
	\affiliation{Department of Physics, Indian Institute of Technology Kanpur, Kalyanpur, Kanpur, Uttar Pradesh 208016, India.}
	\author{Aditya Mehra}
	\email{aditya.mehra@ed.ac.uk}
	\affiliation{School of Mathematics and Maxwell Institute for Mathematical Sciences, University of Edinburgh, Peter Guthrie Tait Road, Edinburgh EH9 3FD, United Kingdom.}
	\author{Augniva Ray}
	\email{augniva.ray@apctp.org}
	\affiliation{Asia Pacific Center for Theoretical Physics, Postech, Pohang 37673, Korea.}
	
	\begin{abstract}
		Carrollian field theories at the classical level possess an infinite number of space-time symmetries, namely the supertranslations. In this article, we inquire whether these symmetries for interacting Carrollian scalar field theory survive in the presence of quantum effects. For interactions polynomial in the field, the answer is in the affirmative. We also study a renormalization group flow particularly tailored to respect the manifest Carroll invariance and analyze the consequences of introducing Carroll-breaking deformations. The renormalization group flow, with perturbative loop-level effects taken into account, indicates a new fixed point apart from the Gaussian ones.
	\end{abstract}
	\maketitle
	%\vfill
	%
	\section{Introduction}
	
	The proposal of celestial holography\cite{Strominger:2013jfa, He:2014laa, Kapec:2014opa, Kapec:2016jld, He:2017fsb, Ball:2019atb, Kapec:2017gsg, Fotopoulos:2019vac, Stieberger:2018onx}, nucleated by ideas from (i) asymptotic symmetries (BMS group) of asymptotically locally flat space-time, and (ii) soft theorems of gluon and graviton scattering amplitudes and 2D conformal field theories (CFTs) has garnered great importance in the past decade. The key ingredient in this direction is the observation that the soft theorems for graviton $S$-matrix elements in asymptotically flat space-time can be considered equivalent to the Ward identities for asymptotic symmetries (large gauge transformations for gluons). On the (celestial) spherical sections of null infinity $\mathcal{I}^+$, the asymptotic symmetry group acts as conformal transformations and supertranslation along the null directions off the sphere. As an example of the success of the above proposal, it was shown that tree-level gluon/graviton\footnote{Note that for gravity, the Mellin transform prescription needs to be modified to regulate UV divergences.} scattering amplitudes in a Mellin-transformed basis have the same form as correlation functions of a 2D CFT \cite{Pasterski:2017kqt, Banerjee:2019prz}. 
	
	In this context, the first caveat to note is that these correlation functions are distribution valued on the celestial sphere. In particular, they are ultralocal in space due to additional constraints of bulk translation invariance \cite{Banerjee:2018gce}. Second, in the earlier studies, the dual CFT was proposed to live on the codimension-2 celestial sphere, and hence the amplitudes are null-time independent. However, the definition of modified Mellin transform (for UV finiteness of graviton amplitudes) makes it more natural to consider \cite{Banerjee:2020kaa} the celestial conformal primary field to have null/retarded time dependence throughout the null infinity. In this setting, the celestial primaries transform under the full BMS group with nontrivial action corresponding to the null time coordinate.
	
	Both the above points strongly indicate at physics on a Carrollian space-time \cite{AIHPA_1965__3_1_1_0}. This is mainly because, from a field theory point of view, it is reasonable to argue that a theory without spatial gradient terms has ultralocal correlations, of course, at the cost of losing Lorentz invariance. This is expected as $\mathcal{I}^+$ on its own is a non-Riemann manifold; rather is a Carrollian one to be precise. Studies on dynamical Carrollian theories have a moderately rich history \cite{Bacry:1968zf, Henneaux:1979vn}. In the past decade, it has been established that conformal Carrollian isometries are isomorphic to the BMS group \cite{Duval:2014uva, Duval:2014lpa, Bagchi:2010zz}. Unlike Lorentz-invariant theories, the global part of the conformal Carrollian isometries does not uniquely fix the propagator. There are a couple of possibilities, with one of them being time dependent. This particular branch of the two-point function is Dirac delta valued and hence ultralocal in space. It is noteworthy that initial investigations into ultralocal quantum field theories date back almost seven decades \cite{Klauder:1970cs}.
	
	The above set of ideas led to a more generic proposal for Carrollian holography \cite{Bagchi:2022emh}, which states that correlation functions in a conformal Carrollian field theory give the scattering amplitudes in asymptotically flat space-time, in the modified Mellin basis. This was further strengthened \cite{bagchi2023ads} by reproducing 3D (including one null time-like direction) conformal Carroll correlation functions by taking the flat space limit of AdS$_4$ Witten diagrams. Another parallel approach of connecting celestial holography with Carroll-symmetric dual field theory via sourced Carroll Ward identities \cite{Donnay:2022wvx, Donnay:2022aba}, and a pure intrinsic-symmetry-based treatment \cite{Saha:2023hsl} were recently developed. It should also be mentioned that for $\left(2+1\right)$-dimensional bulk, many results have been derived using the 2D boundary conformal Carrollian theories. These include thermal and entanglement entropy \cite{Bagchi:2012cy, Barnich:2012xq, Bagchi:2014iea}, progress towards the bootstrap program \cite{Bagchi:2016geg}, and bulk reconstruction \cite{Hartong:2015usd} among others. 
	
	The studies of Carrollian theories in connection with flat-space holography rely on conformal Ward identities for Carroll primaries. Keeping in mind that the Carrollian holographic proposal was motivated by a concrete model of free scalar field theory and its two-point function, it is now natural to inquire how far one can probe the structures of the correlation function beyond the two-point function by incorporating interactions, particularly when quantum effects are considered by going to loop-level calculations\footnote{Of course we cannot aim at reproducing flat space scattering amplitudes with a single scalar of fixed weight because the Mellin amplitudes require primary fields whose weights are parameterized by a real number.}. While working towards this goal in an example of scalar field theory, we came across curious notions regarding Carrollian energy scales. As one approaches the ultrarelativistic limit from a Lorentzian setup $c \rightarrow 0$, along with decoupling of space and time, energy and momenta also get decoupled. Because of the degenerate Carroll metric, there is no notion of plane-wave solutions, and high-energy degrees of freedom do not necessarily describe physics at short wavelengths. Throughout this paper, we study a number of loop-level quantum features of such example \textcolor{black}{interacting} field theories to understand various facets of this idea. \textcolor{black}{The decoupling of high-energy behaviour from short wavelength ones results in drastically different divergence structures in loop calculations. In fact, the high-energy contribution to the loop integrals is finite, similar to the case of single-particle quantum mechanics. On the other hand, spatial ultralocality gives rise to delta-function divergences, which we regulate using spatial lattice cutoff.}
	
	One of the driving forces behind studying Carrollian field theories \cite{basu2018dynamical, Bagchi:2016bcd, bagchi2019field, Banerjee:2020qjj, Islam:2023rnc, Mehra:2023rmm, Bagchi:2019clu} was the tantalizing property of an infinite number of space-time symmetry generators. Hence, a pertinent question along these lines should be: \textit{what happens to the infinite number of supertranslation symmetry generators in the presence of interactions?} In this article, we address this in a couple of ways: (i) by directly verifying the supertranslation Ward identities for higher-point functions at loop level, and (ii) by studying a renormalization group (RG) flow towards the infrared (IR). The latter procedure has been tailor-made to suit Carroll backgrounds. In particular, lowering of the energy is kept decoupled from spatial scaling. This procedure manifestly keeps the supertranslation symmetries non-anomalous.
	
	The article is organized as follows. In section \ref{sec2}, we briefly review the kinematical space-time symmetry generators of a flat Carrollian manifold and how the global parts of these symmetries, including the conformal ones, constrain the two-point functions. In section \ref{sec3}, we introduce the Carrollian scalar action. After briefly reviewing existing results from canonical quantization, we discuss the generic features of the spectrum of such a theory and the effect of ultralocality on the entanglement structure of energy eigenstates. In section \ref{sec4} we carry out the lowest order loop calculation in perturbative quantization of the Carrollian scalar field in three and four dimensions, respectively. The results include renormalization of the mass and supertranslation Ward identities of correlation functions. In section \ref{sec5}, we introduce an RG program suited for scalar Carrollian theory and study RG flow. We find a new fixed point in the small dimensional parameter subspace for the theories in $d=3$ and $d=4$. We conclude in section \ref{conclusions}, alluding to a number of interesting studies worth investigating in the near future.	

	\section{Brief review of conformal Carrollian symmetries}\label{sec2}
	
	The most straightforward way to understand the structure of Carroll (flat) manifold 
	\cite{Duval:2014uoa,Bergshoeff:2014jla,Dutta:2022vkg} is by looking at the ultra-relativistic limit \cite{basu2018dynamical, Banerjee:2020qjj,Bagchi:2019clu,Bagchi:2019xfx,Bagchi:2016bcd} of $d$ dimensional Minkowski space-time
	\begin{equation}\label{stc}
		x^i \rightarrow x^i,\quad t \rightarrow \epsilon t,\quad \epsilon \rightarrow 0 
	\end{equation}
	where $i=1,...,d-1$. This is the same as the $c\rightarrow 0$ limit of \cite{AIHPA_1965__3_1_1_0}. This can equivalently be understood as an In\"{o}n\"{u}-Wigner contraction of the conformal symmetry group of Minkowski space. Operationally, we apply \eqref{stc} on the relativistic conformal generators \cite{{Duval:2014uva}} and regularize them. \textcolor{black}{The generators resulting from this procedure are listed in Table \ref{tab:cvb}}.
	\begin{table}[t]
		\begin{center}
			\begin{tabular}{ |p{.5cm}|p{4cm}|p{7cm}|}
				\hline
				& Transformations & Generators\\
				\hline
				1.  &Translation&$H=\p_t, ~ P_i=\p_i$\\
				2.&  Rotation& $J_{ij}=(x_i \p_j-x_j \p_i)$\\
				3. &Boost&$B_i=x_i \p_t$\\
				4. &Scale transformation&$D= (t \p_t+x_i \p_i)$\\
				5.& Special conformal transformation&$K_j=2x_j(t\p_t+x_i\p_i)-(x_i x_i)\p_j, K= x_i x_i \p_t$\\
				\hline
			\end{tabular}
		\end{center}
		\caption{Conformal Carrollian generators.} \label{tab:cvb}
	\end{table}
	
	These generators form a Lie algebra \cite{Bagchi:2022eui} called finite conformal Carrollian algebra (CCA). The nonvanishing brackets are given by \cite{Chen:2023pqf,Islam:2023rnc}
	\bea{}\label{algebra}
	&&\non [J_{ij}, B_k ]=\delta_{k[j}B_{i]}, ~ [J_{ij}, P_k ]=\delta_{k[j}P_{i]},~ [J_{ij}, K_k ]=\delta_{k[j}K_{i]}, ~ [B_i,P_j]=-\delta_{ij}H,\\
	&&\non  [B_i,K_j]=\delta_{ij}K,~ [D,K]=K,~[K,P_i]=-2B_i,~[K_i,P_j]=-2D\delta_{ij}-2J_{ij},\\ &&[H,K_i]=2B_i,~[D,H]=-H, ~[D,P_i]=-P_i,~[D,K_i]=K_i.
	\eea
	One interesting property of the CCA is that it admits an infinite dimensional extension. The generators for the infinite extended algebra in $d$ dimensions can be written as \cite{Bagchi:2016bcd}
	\begin{equation} \label{sutra}
		M_f =f(x^{1},x^{2},...,x^{d-1})\p_t =:f(x)\p_t.
	\end{equation}
	Here, $f(x)$ are arbitrary tensors and transform under irreducible representations of $so(d - 1)$. If we take $f(x)= (1,x_i,x_kx_k)$ we get the finite generators $M_f = (H,B_i,K)$ respectively. The finite generators with $M_f$ form the infinite-dimensional CCA. The transformations generated by $M_f$ are known as supertranslation transformations. 
	
	The Lie brackets between the finite set and the infinite generators are given by
	\begin{eqnarray} \label{infalgebra}
		&&\non [P_i, M_f] =M_{\p_i f},\quad  [D,M_f] =M_h,~\text{where}~h=x_i \p_i f-f,\\
		&&\non [K_i,M_f]= M_{\tilde{h}},~\text{where}~\tilde{h}=2x_i h-x_k x_k\p_i f,\\
		&&[J_{ij},M_f]= M_g,~\text{where}~{g}=x_{[i}\p_{j]}f,\non\\
		&&[M_f,M_g]=0.
	\end{eqnarray}
	From this point onward, we do not need components of the spatial coordinates explicitly and will use the notation $(t, \mathbf{x})$ for temporal and spatial coordinates.
	
	The finite part of CCA, generated by $H, P_i, B_i, J_{ij}, D, K$ and $K_i$, just like the conformal group of Minkowski space, is sufficient enough to constrain the two and three point functions. However, a couple of distinct solutions to the Ward identities lead to the correlation functions. Specifically for the two-point function, the trivial one \cite{Bagchi:2016bcd} is time independent
	\begin{eqnarray}
		\langle \Phi_{\Delta}(t_1, \mathbf{x}_1) \Phi_{\Delta'}(t_2, \mathbf{x}_2)\rangle \sim \frac{1}{|\mathbf{x}_1-\mathbf{x}_2|^{\Delta+\Delta'}} \delta_{\Delta,\Delta'},
	\end{eqnarray}
	where $\Delta$ are the conformal dimensions of these spin-less fields (for simplicity, we did not include spin). On the other hand, there's another, time-dependent solution \cite{Chen:2021xkw,Bagchi:2022emh,bagchi2023ads}, namely the `delta function branch', which takes the form
	\begin{eqnarray} \label{del_branch}
		\langle \Phi_{\Delta}(t_1, \mathbf{x}_1) \Phi_{\Delta'}(t_2, \mathbf{x}_2)\rangle \sim \frac{\delta^{d-1}(\mathbf{x}_1-\mathbf{x}_2)}{(t_1-t_2)^{\Delta+\Delta' - d+1}}.
	\end{eqnarray}
	The realization of this delta function branch \footnote{\textcolor{black}{One way to understand the rotational invariance of \eqref{del_branch} is by smearing the delta function with two arbitrary test functions: $\int d^{d-1}\mbx_1\, d^{d-1}\mbx_2\,  \delta^{d-1}(\mbx_1 -\mbx_2) f(\mbx_1) g(\mbx_2) = \int d^{d-1}\mbx \, f(\mbx) g(\mbx)$. The final answer remains the same whether $\mathbf{x}$ or $\mathbf{Rx}$, $\left(\mathbf{R} \in O\left(d-1\right) \right)$ is used as the dummy integration variable.}} was corroborated using the example of a free scalar field of mass $m \rightarrow 0$ in \cite{Bagchi:2022emh}.
	%%%%%
	\section{Spectrum of Carrollian quantum field theory} \label{sec3}
	
	In this section, we start by reviewing some of the results in \cite{Bagchi:2022emh} in reference to canonical quantization of the simple free Carrollian scalar in arbitrary spatial dimensions. Then we move on to extract a more thorough structure of the Hilbert space, for both the free and interacting theories. 
	
	A field that transforms trivially under the action of rotation generators $J_{ij} = x_{[i} \p_{j]}$ will be called a scalar. The most straightforward theory of an interacting scalar in a $d$ dimensional flat Carroll manifold is of the form
	\begin{eqnarray} \label{generic_action}
		S = \int dt\, d^{d-1}\mathbf{x} \left(\frac{1}{2}\left(\partial_t\phi \right)^2 - V(\phi) \right).
	\end{eqnarray}
	The supertranslation transformations \eqref{sutra} act on the scalar field as 
	\begin{eqnarray} \label{sutra_phi}
		\phi \rightarrow \phi + f(\mathbf{x}) \p_t \phi . 
	\end{eqnarray}
	The absence of a spatial derivative in the Lagrangian makes the theory manifestly invariant under supertranslation transformations. For classical aspects of scalars on generic Carroll background, see \cite{Baiguera:2022lsw}.
	
	Notably, although space-time democracy is lost in the inherent geometric structure of Carroll space-time, the dilatation generating vector field retains the same form $D = t\p_t + x^i \p_i$ as that of Minkowski space-time, as shown in Table \ref{tab:cvb}. This sets the naive scaling dimension $\Delta_{\phi} = \frac{d-2}{2}$ \footnote{However, we will see later in section 5, that there is a separate idea of dimensions of operators \textcolor{black}{that} becomes more \textcolor{black}{ appropriate} for Carrollian theories.}. The generic theory is therefore classically symmetric under the scaling and special conformal transformations ($K, K_i$) only for $V(\phi) \sim \phi^6$ in $d=3$ and for $\phi^4$ in $d=4$.
	
	We do not attempt to quantize the classically conformally invariant theory discussed above but rather introduce a mass term. The reasons are two-fold. First, as there are no spatial derivatives, the free theory Lagrangian consists of just the temporal derivative term $\sim \frac{1}{2} (\p_t \phi)^2$ and can be thought of as an (uncountably infinite) bunch of free particles each designated by a point in the real space. As free particle energy eigenstates are not well defined, keeping a mass term is a good idea so that the system is now replaced by a bunch of harmonic oscillators already decoupled in real space. The second rationale comes from the Wilsonian renormalization group point of view. In a Lorentz invariant scalar theory, even if one starts from a massless theory at some scale, as one flows towards the IR, it gains mass in the presence of coupling (we are concerned about marginal ones here), breaking scale invariance. As we will see later in the paper, this feature is also true in our Carrollian scalar. Also, a massive theory on a compact spatial manifold, e.g. a $(d-1)$ sphere, can still be conformally invariant when the mass is identified with the inverse radius of the sphere. However, we do not attempt to do that and keep the spatial topology as $\mathbb{R}^{d-1}$. For recent developments on Carroll scalars on curved manifolds, please refer to \cite{Rivera-Betancour:2022lkc}.
	%%%%%
	\subsection{Spectrum of the free theory}
	
	Let us start with the free massive scalar field on a flat Carrollian manifold
	\begin{equation} \label{action_free_d} 
		S=\int dt\, d^{d-1}\mathbf{x} \left[\frac{1}{2}\left(\partial_t\phi \right)^2 - \frac{1}{2} m^2 \phi^2 \right].
	\end{equation}
	This leads to the Euler-Lagrange e.o.m.
	\begin{equation} \label{eom_free}
		\left(\partial_t^2 + m^2\right)\phi\left(t, \mathbf{x}\right) = 0.
	\end{equation}
	It is easily seen that the Heisenberg picture scalar field operator can be expressed in terms of creation-annihilation fields
	\begin{equation} \label{phiaadag}
		\phi(t, \mathbf{x})=\frac{1}{\sqrt{m}}\left[a^{\dag}\left(\mathbf{x}\right) e^{i m t} + a\left(\mathbf{x}\right) e^{-i m t} \right],
	\end{equation}
	which satisfy the canonical quantization condition
	\begin{equation}\label{eq:commutationreln}
		\left[a\left(\mathbf{x}\right), a^{\dag}\left(\mathbf{x'}\right)\right] = \frac{1}{2} \delta^{\left(d-1\right)}\left(\mathbf{x}-\mathbf{x'}\right).
	\end{equation}
	In contrast with the usual Lorentzian quantum field theory (QFT), the creation and annihilation operators are defined here in real space-time. Finally, the normal ordered Hamiltonian is given by
	\begin{equation} \label{hamiltonian}
		H= 2 m\int d^{d-1}\mathbf{x} \,%\left[2
		a^{\dag}\left(\mathbf{x}\right) a\left(\mathbf{x} \right).% + \frac{1}{2}\delta^{\left(d-1\right)}\left(0 \right)\right],
	\end{equation}
	Without normal ordering, one gets an irrelevant but infinitely diverging additive piece $\int \delta^{\left(d-1 \right)}(0)$ which can be ignored. The vacuum state $|0\rangle$, satisfying $H|0\rangle = 0$ gives rise to
	\begin{equation} \label{ground_st}
		a\left(\mathbf{x}\right) \ket{0}  = 0, \quad \forall \mathbf{x}\, .
	\end{equation}
	The first quantum excitation is
	\begin{equation}
		\ket{\mathbf{x}} = a^{\dag}\left(\mathbf{x}\right)\ket{0} = m\ket{0},
	\end{equation}
	and has energy $m$, which is irrespective of $\mathbf{x}$, a direct consequence of the fact that the Hamiltonian \eqref{hamiltonian} has trivial dispersion\footnote{Dispersionless free Hamiltonians are extremely important from the perspective of present-day understanding of correlated electrons, observed in Moire pattern condensed matter systems \cite{Tarnopolsky:2018mxs}.}. As opposed to particle interpretation in relativistic QFTs, we see the quantum fluctuations are synchronized throughout space, as all such excitations have the same energy. The time ordered two-point function\footnote{Similar contact term behaviour has been recently reported for propagators in Carrollian gauge theories as well \cite{Islam:2023rnc}.} follows straightaway from the quantization condition and the definition of the vacuum state as
	\begin{eqnarray} \label{2pt_first}
		\langle 0|T\left(\phi\left(t_1, \mathbf{x}_1\right) \phi\left(t_2, \mathbf{x}_2\right)\right) |0\rangle = \frac{1}{2m} e^{-im|t_1 - t_2|} \delta^{d-1} \left(\mathbf{x}_1 - \mathbf{x}_2\right).
	\end{eqnarray}
	Of course, it diverges as $m\rightarrow 0$, which is expected because then the system is just a bunch of free particles whose wavefunctions are not normalizable. But with the mass term present, we can systematically track the divergences. For small $m$, we get from \eqref{2pt_first}, the two-point function $\sim \left(\frac{1}{2m} -\frac{i}{2} |t| \right) \delta^{d-1} \left(\mathbf{x_1} - \mathbf{x_2}\right)$, where $t = \left( t_2 - t_1 \right)$ is the temporal separation between the two fields. The time dependent term matches exactly, with \eqref{del_branch}, keeping in mind that $\Delta = \frac{d-2}{2}$. This can be compared with the apparent vanishing or blowing up of the normalization constants in two-point functions in Carrollian theories viewed from the celestial perspective \cite{bagchi2023ads}. There, finiteness \cite{Salzer:2023jqv} requires a subtle adjustment between the mass of the celestial massive operators, the dimensions $\Delta$ and the AdS radius $R$, which would eventually be set to zero for asymptotically flat space-time, whose duals are expected to be Carrollian theories.
	%%%
	\begin{figure}[t]
		\centering
		\includegraphics[scale=0.25]{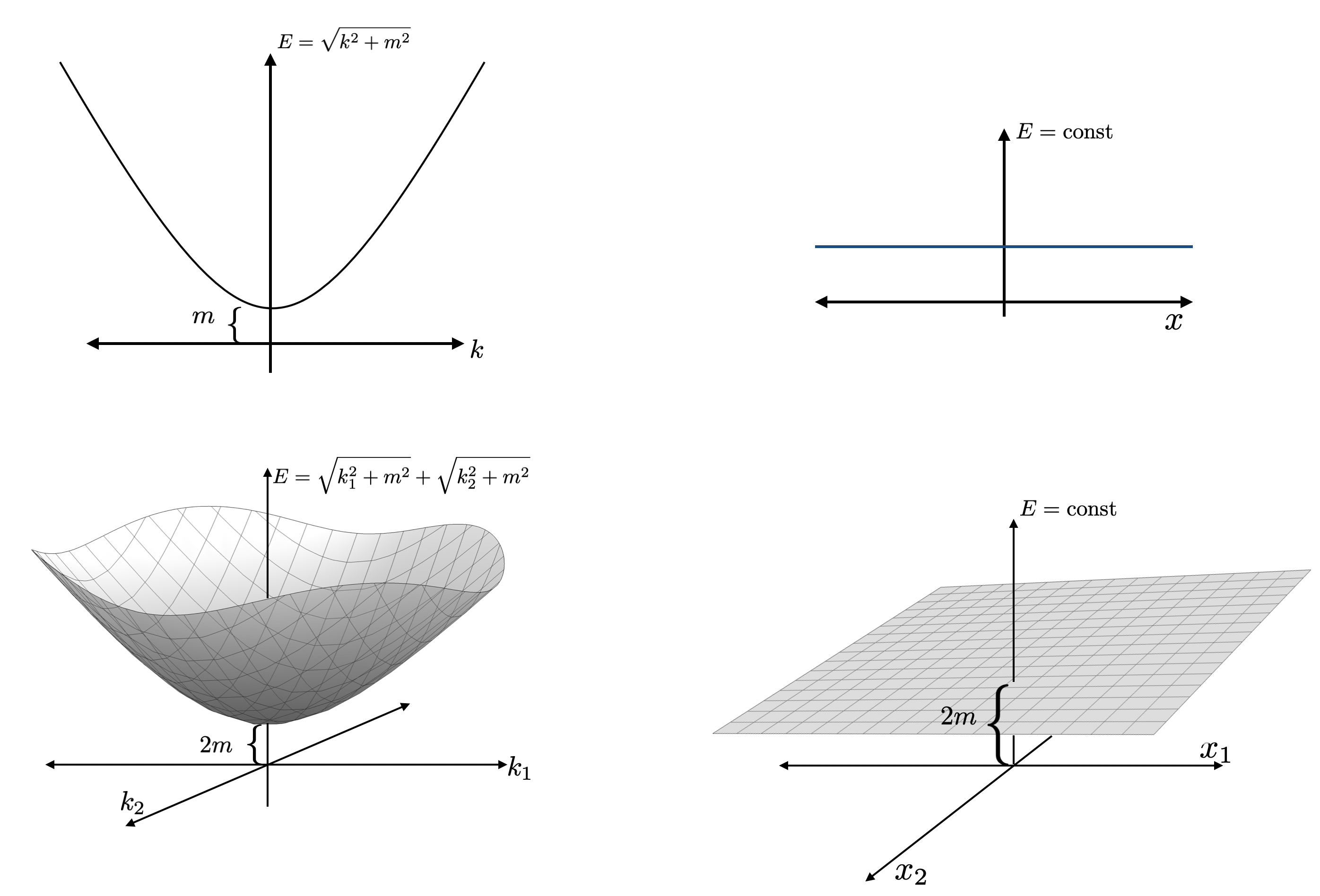}
		\caption{Left: single particle $a^{\dagger}_k \ket{0}$ and two particle spectra $a^{\dagger}_{k_1} a^{\dagger}_{k_2} \ket{0}$ of a relativistic massive free theory in $1+1$ dimension. Right: degeneracy in a massive Carrollian theory for the single $a^{\dagger}(x) \ket{0}$ and double excitations $a^{\dagger}(x_1) a^{\dagger}(x_2) \ket{0}$}
		\label{fig:my_label}
	\end{figure}
	%%%
	Higher excited states are obtained by repeated application of $a^{\dag}\left(\mathbf{x}\right)$, e.g. consider the state
	\begin{equation} \label{N_eigenstate}
		\ket{\mathbf{x}_{12\ldots N}} := a^{\dag}\left(\mathbf{x}_N\right) a^{\dag}\left(\mathbf{x}_{N-1}\right)\ldots a^{\dag}\left(\mathbf{x}_1\right) \ket{0} = \prod_{\mathrm{j} = 1}^{N} a^{\dag}\left(\mathbf{x}_j\right) \ket{0},
	\end{equation}
	analogous to an $N$ particle state of relativistic QFT. In fact, these excited states are also infinitely degenerate
	\begin{equation}
		H\ket{\mathbf{x}_{12\ldots N}} = N\,m\ket{\mathbf{x}_{12\ldots N}}.
	\end{equation}
	(see Fig. \ref{fig:my_label} for an illustration of the spectra in relativistic and Carrollian theories.) Each energy level being infinitely degenerate is a direct consequence of the infinite number of conserved charges associated with supertranslation symmetries. Another obvious but noteworthy point to mention is that at the limit $m \rightarrow 0$, all the states collapse to the ground state of zero energy due to the restoration of scaling symmetry. A similar observation has been made for tensionless strings \cite{Bagchi:2015nca, Bagchi:2019cay}, where all perturbative closed string excitations collapse to the induced vacuum of the open string. This is interesting and more than a mere coincidence because the tensionless string theory worldsheet enjoys Carrollian symmetry in a conformal gauge. Just after the appearance of the first version of the present article another interesting piece of work \cite{deBoer:2023fnj} appeared, which emphasized that the infinite degeneracy poses a problem in defining the canonical partition function and one way to bypass that might be via going to a grand canonical ensemble.
	%%%
	\subsubsection*{Degeneracies in the interacting theory}
	
	As argued earlier, any action of the form \eqref{generic_action} is invariant under Carrollian supertranslation symmetries $\delta \phi = f(\mathbf{x})\p_t \phi $, including the Carrollian boost. Hence, we choose interaction $V(\phi) = \lambda \phi^n$ for simplicity.
	
	Although the full spectrum of the above theory could, in principle, be found perturbatively, one can make a few qualitative remarks immediately based on simple reasoning without getting into the details.
	
	We recall that the free Carrollian scalar field theory is a collection of linear harmonic oscillators localized at each point in real space, completely decoupled from one another, all having the same frequency. Since the perturbation $\sim \int \phi^n$ does not violate ultralocality, the degeneracies found in the free theory are intact. More concretely, under perturbation, the single excitation free theory state $\ket{\mathbf{x}} \rightarrow \ket{\mathbf{x}}'$ has the same degeneracy, i. e. same energy irrespective of the point $x$.
	%%%
	\subsection{Entanglement entropy}
	Measurement of quantum entanglement between the degrees of freedom between two sub-regions of a quantum field theory is best captured by the entanglement entropy. This, for free relativistic theories in the ground state, has been calculated in \cite{Bombelli:1986rw, Calabrese:2004eu, Casini:2009sr, Hertzberg:2012mn}.

    \textcolor{black}{A direct consequence of the ultralocality or absence of any coupling between neighbouring degrees of freedom in the concerned Carrollian theory is the vanishing of the sub-region entanglement entropy (here, we only focus on subregions at equal time, unlike the more generalized version of the covariantized entanglement entropy } \cite{Hubeny:2007xt})\textcolor{black}{ in the ground state.} The easiest way to see this is that the ground state \eqref{ground_st} is a tensor product of the ground states of all the local oscillators \eqref{phiaadag} sitting at each point of space. Hence, the ground state density matrix $\rho = |0 \rangle \langle 0|$ gives us zero von-Neumann entropy between any two subregions of space at a given time. The same conclusion holds true for the free Carrollian fermions discussed in \cite{Bagchi:2022eui}, whose action is of the form 
	\begin{eqnarray}
		S_F \sim \int d^d x \,\bar{\psi} \Gamma \p_t \psi.
	\end{eqnarray}
	However, there are free Carrollian theories (not considering Carrollian gauge theories \cite{Bagchi:2016bcd, basu2018dynamical,Islam:2023rnc, Bagchi:2019clu, Bagchi:2019xfx, banerjee2021interacting, deBoer:2021jej, Bidussi:2021nmp}), at least for fermions, whose actions may contain spatial derivatives as well. One such prominent example is the dynamics of electrons in bilayer graphene Moire pattern at `magic angles', as expounded in \cite{Bagchi:2022eui}. Just like the case of the free ultralocal theory described above, this also is dispersionless and has Carrollian symmetry. However, there are neighbourhood site couplings, bringing in non-zero entanglement entropy between two sub-regions.
	%%%
	\subsubsection*{Entanglement entropy for $d=2$}
	The case of 2-dimensional Carrollian theories requires special discussion. The symmetry group, including conformal transformation, is generated by the BMS$_3$ algebra \cite{Brown:1986nw, Bagchi:2010zz, Bagchi:2012cy}
	\begin{eqnarray}
		&&  [L_n,L_m] = (n-m) L_{m+n} + \frac{c_L}{12} (n^3 - n) \delta_{m+n,0} \, , \nonumber\\
		&& [L_n,M_m] = (n-m) M_{m+n} + \frac{c_M}{12} (n^3 - n) \delta_{m+n,0} \, , \quad [M_m, M_n] =0.
	\end{eqnarray}
	This also serves as the asymptotic symmetry group of asymptotically flat space-times in 3 bulk dimensions. Hence it is natural that $d=2$ conformal Carrollian theories are putative dual candidates to gravity in asymptotically flat space-time. The proposal of this duality has been substantiated via a large number of checks \cite{Barnich:2012xq, Bagchi:2012xr, Bagchi:2012yk,Bagchi:2013qva, Bagchi:2014iea}. One such check is sub-region entanglement entropy in the dual field theory. For an 1 dimensional sub-region of spatial extent $l_{x}$ and temporal extent $l_t$, the entanglement entropy is given by
	\begin{eqnarray}
		S = \frac{c_L}{6} \ln \left( \frac{l_x}{a}\right) + \frac{c_{\textcolor{black}{M}}}{6} \left( \frac{l_t}{l_x}\right).
	\end{eqnarray}
	For Einstein gravity in the bulk, $c_L = 0$ and $c_M = \frac{3}{G}$ ($G$ is the Newton's constant for 3d gravity). The answer for sub-region entanglement entropy for a purely spatial sub-region $l_t = 0$, for a Carrollian theory dual to Einstein gravity must be equal to zero. 
	
	On the other hand, the arguments we placed for vanishing sub-region entanglement entropy for ultralocal Carroll theories would hold for arbitrary $m$ and even for $m=0$, i.e. for classically conformal field theories. Hence, we can claim that $d=2$ Carrollian theories with ultralocal behaviour are feasible candidates for a theory dual to Einstein gravity in asymptotically flat space-time in 3 dimensional bulk.
	%%%%%
	\section{Perturbative quantization} \label{sec4}
	
	\subsection{The lattice regularization}\label{lattice}
	
	For the purpose of perturbative quantization, we start with the lattice regularized version of \eqref{generic_action}. For the time being, we focus on the $d=3$ case, where the marginal deformation is $\phi^6$. Keeping this in mind, the lattice discretized Hamiltonian takes the form (there are no neighbour hopping terms due to ultralocalization of \eqref{generic_action})
	\begin{eqnarray} \label{HJ}
		H = \sum_i \frac{P_i^2}{2 M } + \frac{1}{2} K X_i^2 + \frac{\tilde\Lambda}{6!} X^6_i,
	\end{eqnarray}
	where $\tilde\Lambda$ is the coupling with mass dimension 7. %$\left[\tilde\Lambda\right] = \left[M L^{-4} T^{-2}\right]$.
	The discretized theory is equipped with the usual canonical commutation relations.
	
	The time-ordered\footnote{All correlation functions appearing below are time-ordered. Hence $\langle T(X(t_1) X(t_2) )\rangle$ is written as $\langle X(t_1) X(t_2) \rangle$} two-point function in the vacuum of the perturbed Hamiltonian \eqref{HJ} is
	\begin{equation} \label{2pt_corrected_Euclidean}
		\langle X(t) X(0)\rangle = \sum_{n>0} \left| \langle \tilde{0}|X|\tilde{n}\rangle \right|^2 e^{-i\left(\tilde{E}_{n} - \tilde{E}_{0}\right)|t|},
	\end{equation}
	where $\ket{\tilde{n}}$ are the eigenstates and $\tilde{E}_n$ are the eigenvalues of the interacting Hamiltonian \eqref{HJ}.
		
	Up to the first order in $\tilde{\Lambda}$, the sum in the above correlation function receives contribution only from $|{\tilde{n}}\rangle = |{\tilde{1}}\rangle$, and the expression is given by
	\begin{equation} \label{correlator_discrete_phi6}
		\langle X_j(t) X_k(0)\rangle = \frac{\hbar}{2 M \omega}\left(1 - \frac{\tilde\Lambda}{64} \frac{\hbar^2}{M^3\omega^4} \textcolor{black}{+\mathcal{O}\left(\frac{\tilde{\Lambda}^2 \hbar^4}{M^6 \omega^6}\right)}\right) e^{-i\omega\left(1 + \frac{\tilde\Lambda}{64} \frac{\hbar^2}{ M^3\omega^4}\textcolor{black}{+\mathcal{O}\left(\frac{\tilde{\Lambda}^2 \hbar^4}{M^6 \omega^6}\right)} \right)|t|} \delta_{jk},
	\end{equation}
	where $\omega = \sqrt{K/M}$. Here \textcolor{black}{$\mathcal{O}\left(\frac{\tilde{\Lambda}^2 \hbar^4}{M^6 \omega^6}\right)$ denotes perturbative corrections beyond first order}. Refer to the Appendix \ref{Proof_2pt} for the steps leading to \eqref{correlator_discrete_phi6}. Assuming that we have defined our Hamiltonian \eqref{HJ} on a square lattice with lattice parameter $a$, the continuum version  can be reached by the following substitutions
	\begin{eqnarray}\label{4}
		P_i/(\sqrt{M}\, a) \rightarrow \pi(x) ,~~ X_i \sqrt{M}/a \rightarrow \phi(x),~~ \sqrt{\frac{K}{M}} \rightarrow m, ~~ \frac{a^4 \tilde\Lambda}{M^3} \rightarrow \tilde\lambda.
	\end{eqnarray}
	Finally, taking $a \rightarrow 0$ yields the following continuum Hamiltonian
	\begin{eqnarray} \label{HJ1}
		H=\int d^2x \left(\frac{1}{2}\pi^{2}+\frac{1}{2}m^2\phi^2+\frac{\tilde\lambda}{6!}\phi^6\right),
	\end{eqnarray}
	where $\pi=\dot{\phi}$. With the scaling rules \eqref{4}, the two-point function \eqref{correlator_discrete_phi6}  takes the following form with explicit cutoff dependence
	\begin{equation}
		\langle \phi\left(t, \mathbf{x}\right) \phi\left(0, \mathbf{x'}\right) \rangle = \frac{1}{2 m}\left(1 -  \frac{\tilde\lambda}{64m^4 a^4} \right) e^{-im \left(1 + \frac{\tilde\lambda}{64m^4 a^4} \right)| t|} \delta^2\left(\mathbf{x}-\mathbf{x'}\right),
	\end{equation}
	in units where $\hbar=1$. Defining the effective mass as $m_R = m \left(1 + \frac{\tilde\lambda}{64m^4 a^4} \right)$, we find that 
	\begin{equation} \label{2ptcont}
		\langle \phi\left(t, \mathbf{x}\right) \phi\left(0, \mathbf{x'}\right) \rangle = \frac{1}{2 m_R} e^{-im_R |t|} \delta^2\left(\mathbf{x}-\mathbf{x'}\right),
	\end{equation}
	is the 1-PI effective two-point function. \textcolor{black}{We emphasize that this is correct only up to first order in $\tilde{\lambda}$. The next to leading order corrections are of the form $m_R= m \left(1 + \frac{\tilde\lambda}{64m^4 a^4} + \mathcal{O}\left(\frac{\tilde{\lambda}^2 }{m^8 a^8}\right)\right)$, which follows from the form of the higher order terms appearing in \eqref{correlator_discrete_phi6}}. \textcolor{black}{Equation \eqref{2ptcont}} has the same behaviour as reported in \cite{Bagchi:2022emh}, where  the mass has now been renormalized. The renormalization is, of course, divergent as one goes to the continuum limit and can be absorbed by introducing a counter term as is routine in \textcolor{black}{Quantum Field Theories} (QFTs).
	
	The above analysis for a bunch of anharmonic oscillators on a 2D lattice was designed to capture the physics of a $d=3$ Carroll scalar. By following similar routes, we get equivalent results for a $d=4$ theory with $\phi^4$ coupling. The discretized Hamiltonian, in this case, is
	\begin{equation} \label{HJ_four}
		H = \sum_i \frac{P_i^2}{2 M } + \frac{1}{2} K X_i^2 + \frac{\Lambda}{4!} X^4_i,
	\end{equation}
	with the mass dimension of the coupling constant $\Lambda$ being 5. Here, the two-point function for $t>0$, up to first order in perturbation theory (in $\Lambda$) is\footnote{This can be compared with the classical perturbation analysis of a quartic anharmonic oscillator, where one needs to `renormalize' the natural frequency to get a bounded solution. Consequently, it matches exactly with the classical result \textcolor{black}{(see for example \cite{jose2000classical}, section 6.3 or \cite{landau1976mechanics} section 29)} once one replaces the amplitude by $\sqrt{\langle X^2(0)\rangle}$, calculated in the free vacuum.}
	\begin{equation} \label{tpf}
		\langle X_j(t) X_k(0)\rangle = \frac{\hbar}{2 M \omega}\left(1 - \frac{\Lambda}{4!} \frac{3\hbar}{M^2\omega^3}\right) e^{-i\omega\left(1 + \frac{\Lambda}{4!} \frac{3\hbar}{M^2\omega^3} \right)|t|} \delta_{jk}.
	\end{equation}
	With proper dimensions, the equivalent of the continuum limit prescription \eqref{4} for this case would be
	\begin{eqnarray}\label{5}
		\frac{P_i}{\sqrt{Ma^3}} \rightarrow \pi(x) ,~~ X_i \sqrt{\frac{M}{a^3}} \rightarrow \phi(x),~~ \sqrt{\frac{K}{M}} \rightarrow m, ~~ \frac{a^3 \Lambda}{M^2} \rightarrow \lambda.
	\end{eqnarray} 
	The Hamiltonian in $a\rightarrow0$ limit becomes
	\begin{eqnarray} \label{HJ2}
		H=\int d^3x \left(\frac{1}{2}\pi^{2}+\frac{1}{2}m^2\phi^2+\frac{\lambda}{4!}\phi^4\right).
	\end{eqnarray}
	The continuum version of the two-point function\eqref{tpf} takes exactly the same form of \eqref{2ptcont}, except now that we have the three-dimensional Dirac delta function \textcolor{black}{, and the effective mass is} 
	\begin{equation} \label{mr2}
		m_R = m \left(1 + \frac{\lambda}{8 m^3 a^3} + \mathcal{O}\left(\frac{\lambda^2}{ m^6 a^6} \right) \right) . 
	\end{equation}
	A \textcolor{black}{ comment} regarding the effective mass formula\textcolor{black}{s} \eqref{2ptcont} and \eqref{mr2} is in order. Note that we could produce these results starting from a single particle quantum mechanical calculation only because the field theory is ultralocal in nature and the frequency renormalizations observed in \eqref{correlator_discrete_phi6}, \eqref{tpf} are finite. We did not face the usual divergences faced in relativistic QFTs while performing the equivalent of the one-loop calculation above because the free version of the theory \eqref{HJ1} is dispersionless, and all the oscillator modes have the same ground state energy. \textcolor{black}{In fact higher order perturbative corrections, as suggested in \eqref{correlator_discrete_phi6} and \eqref{mr2}, and also expected from the quantum mechanics of anharmonic oscillators (see, for example \cite{Giorgini:2021esl} and references therein), do not produce any divergence from loop integrals.} However, the divergence is manifest only in a small length scale $a$. Hence, it would be apt to say that this Carrollian theory is super-renormalized in the true ultra-violet (no divergence coming from high frequencies), but is plagued with small-wavelength divergence. In the absence of nontrivial dispersion, these two behaviours are decoupled.
	%%%
	%%%
	\subsection{The diagrammatic approach and higher point functions}
	
	In this section, we study the Carrollian $\phi^6$ theory and $\phi^4$ theory. These are classically invariant under the generators of Carrollian algebra (not under CCA owing to the mass term). The main goal of this section is to get the two-point correlation function in the interacting Carrollian scalar field theory (both $\phi^4$ and $\phi^6$) with the possible one-loop corrections.
	%%%
	\subsubsection{The two-point function revisited}
	
	Before moving to higher point functions, we would first establish that the traditional diagrammatic approach in continuum field theory, when applied to the interacting Carrollian scalars, reproduces the same two-point functions obtained in the last section from the many-body quantum mechanical system.\\ 
	
	\noindent \textbf{\underline{The $d=4$ case}:}
	
	To this end, we first consider the $d=4$ Carroll scalar with $\phi^4$ coupling
	\begin{equation}\label{dfs}
		S=\int dt\, d^{3}\mathbf{x} \,\left[\frac{1}{2}(\partial_t\phi )^2 - \frac{1}{2} m^2 \phi^2  -\frac{\lambda}{4!} \phi^4 \right].
	\end{equation}
	The Feynman propagator for the free Carrollian theory $\left(\lambda = 0\right)$ is found by inverting the operator $\partial_t^2+m^2$ with an $\epsilon$ prescription, and is given by
	\bea{}\label{ffh} \Delta_F (t,\mathbf{x}) = \frac{1}{(2\pi)^4} \int d^3 p\, d\omega ~\frac{e^{i(\mathbf{p.x}-\omega t)}}{(\omega^2-m^2+i\epsilon)} =
	-\frac{i}{2m}e^{-im|t|}\delta^{3}(\mathbf{x}).
	\eea
	This exactly reproduces \eqref{2ptcont} for ${\lambda} = 0$, as was also shown in \cite{Bagchi:2022emh}. A caveat regarding the definition of the Feynman propagator, which serves as the time-ordered two-point function comes via the integration contour or $\epsilon$ prescription. If one is not strict regarding the time-ordered correlator and rather concentrates
  on the principal value of the integration $\int d\omega ~\frac{e^{-i\omega t}}{(\omega^2-m^2)}$, the result is $\sim \frac{\sin{(m\,t)}}{m}\delta^{3}(\mathbf{x})$. This, unlike the time ordered two-point function, has a well defined limit as $m \rightarrow 0$.
	
\begin{figure}[h]
		%    \centering
		%    \begin{subfigure}{0.05\textwidth}
		\centering
		\includegraphics[scale=0.1]{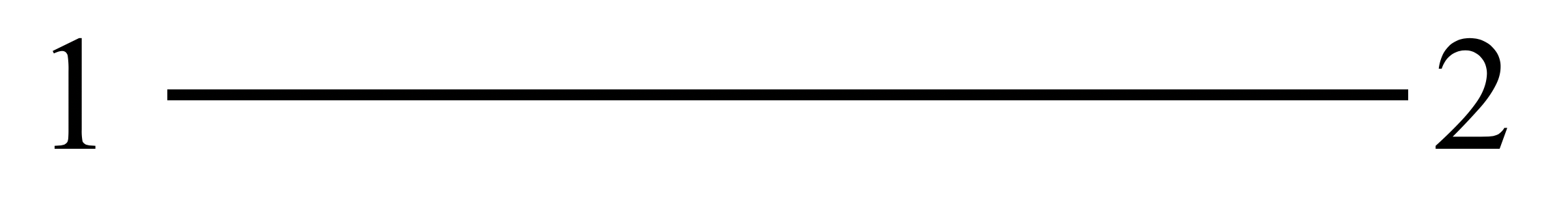}
		%    \end{subfigure}
		\caption{two-point correlation function}
		\label{fig_2pt_1}
	\end{figure}

	We denote the propagator \eqref{ffh} diagrammatically as in Fig. \ref{fig_2pt_1}
 
\begin{figure}[h]
		%    \centering
		%    \begin{subfigure}{0.47\textwidth}
		\centering
		\includegraphics[scale=0.1]{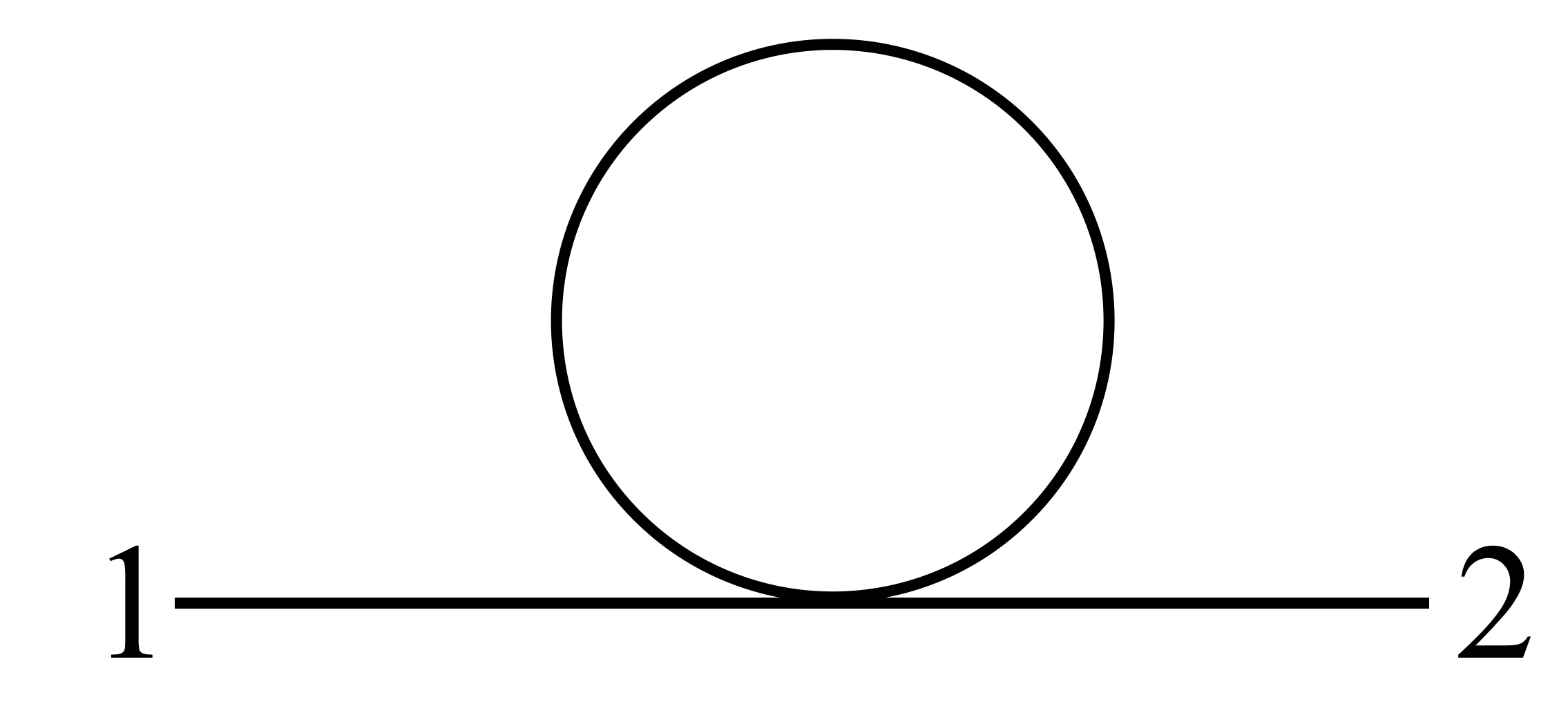}
		%    \end{subfigure}
		\caption{Correction to two-point function in $\phi^4$ theory}
		\label{fig_tadpole}
	\end{figure}

	Let us observe the effect of interaction on the 1-PI propagator. At linear order in $\lambda$, the correction to the two-point function is diagrammatically represented by the connected Feynman graph in Fig. \ref{fig_tadpole}. The contribution of this real space diagram is
	\bea{}
	\deltabar \langle \phi(x_1)\phi(x_2)\rangle = \frac{\lambda}{2}\Delta_{F}(0) \int d^4z ~\Delta_F(z-x_2)\Delta_F(z-x_1).
	\eea
	%
	%The connected diagram corresponding to first order perturbative correction in $\lambda$ to the above is
	%
	Clearly the $\Delta_F(0)$ is divergent as $\delta^3(0)$, as can be seen from \eqref{ffh}. However, at this point, we regulate this by defining an upper cutoff in the spatial momentum integral of \eqref{ffh} and hence introducing a small length cutoff $a$. This is the same spatial cutoff we used for the lattice, at each point of which we kept uncoupled anharmonic oscillators in the previous section. Hence, from now on, we read 
	\bea{} \label{feynzero}
	\Delta_F(0)=-\frac{i}{2m a^3}.\eea
    Combining the free propagator and the \textcolor{black}{1-} loop diagram \ref{fig_tadpole} above for the 1-PI, we get the propagator corrected to first order in perturbation theory as
	\bea{} \label{tres}
	\langle \phi\left(x_1\right) \phi\left(x_2\right) \rangle_{{\lambda}} = \frac{i}{(2\pi)^3}\int \frac{e^{-i{p.x}}}{(\omega^2-m^2+i\epsilon)} \left(1+\frac{\frac{i}{2}{\lambda}\Delta_F(0)}{(\omega^2-m^2+i\epsilon)}\right) d^3 p \textcolor{black}{d \omega}.
	\eea
	For small ${\lambda}$, the term in the bracket above can be written as
	$\left(1-\frac{\frac{i}{2}{\lambda}\Delta_F(0)}{(\omega^2-m^2+i\epsilon)}\right)^{-1}$. Therefore, the above becomes
	\begin{eqnarray} \label{tres1}
		\langle \phi\left(x_1\right) \phi\left(x_2\right) \rangle_{{\lambda}} = \frac{i}{(2\pi)^3}\int \frac{e^{-i{p.x}}}{(\omega^2-m^2-\frac{i}{2}{\lambda}\Delta_F(0)+i\epsilon)}d^3 p \textcolor{black}{d \omega}.
	\end{eqnarray}
	We see that $\langle \phi\left(x_1\right) \phi\left(x_2\right) \rangle_{{\lambda}}$ possesses a pole at
	\begin{eqnarray} \label{newm} 
		\w^2=m^2+\frac{i}{2}{\lambda}\Delta_F(0)=m^2_R\,,
	\end{eqnarray}
	where $m_R$ is the one-loop corrected mass, matching exactly with the one in \eqref{mr2}, upon using the prescription \eqref{feynzero}. This results in, up to the first order in $\lambda$ perturbation,
	\bea{} \label{2ptnew4d}\langle \phi(x_1)\phi(x_2)\rangle_{\lambda} = \frac{1}{2m_R}\delta^{3}(\mathbf{x}_1 -\mathbf{x}_2 ) e^{-i m_R|t_1 -t_2|},
	\eea
	with $m_R$ matching exactly with the expression \eqref{mr2}.
	
	The Feynman propagator, i.e. the free theory time ordered two-point function, is found by inverting the manifest supertranslation invariant operator $\partial_t^2+m^2$. The one-loop renormalized two-point function \eqref{2ptnew4d} has the same form as the free theory with the mass renormalized. Hence it satisfies the supertranslation Ward identities.\\
	
	\noindent \textbf{\underline{The $d=3$ case}:}
	
	Very similar line of analysis carries on for the $\phi^6$ theory in $d=3$:\begin{equation}\label{dfs3}
		S=\int dt\, d^{2}x \,\Big[\frac{1}{2}(\partial_t\phi )^2 - \frac{1}{2} m^2 \phi^2  -\frac{\tilde\lambda}{6!} \phi^6 \Big].
	\end{equation}
	Exactly as before one gets the Feynman propagator as:
	\begin{eqnarray}\label{ffd}
		\langle \phi\left(t_1, \mathbf{x}_1\right) \phi\left(t_2, \mathbf{x}_2\right) \rangle = i \Delta_{F} \left(t_1 -t_2, \mathbf{x}_1-\mathbf{x}_2\right) &=& \frac{1}{2 m} e^{-im |t_1-t_2|} \delta^2\left(\mathbf{x}_1-\mathbf{x}_2\right)
	\end{eqnarray}
	The first order in $\tilde{\lambda}$ correction to the two-point function is two-loop, given by the diagram in Fig. \ref{fig 3} below:
	\begin{figure}[h]
		%    \centering
		%    \begin{subfigure}{0.3\textwidth}
		\centering        \includegraphics[scale=0.1]{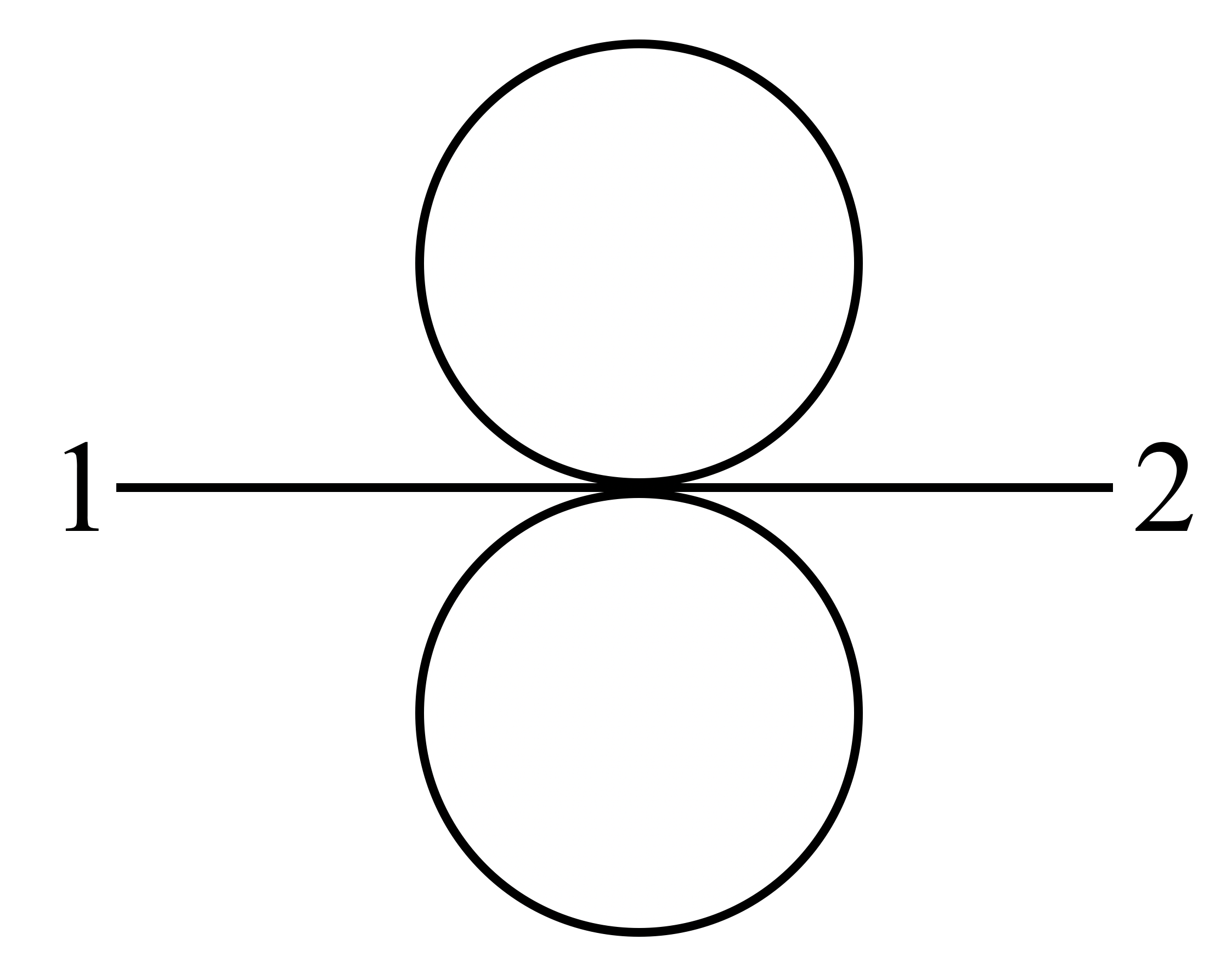}
		%    \end{subfigure}
		\caption{Correction to two-point function in $\phi^6$ theory}
		\label{fig 3}
	\end{figure}

 \noindent In real space this evaluates to the following two-loop integral
	\bea{}\label{kjk}
	%\deltabar \langle \phi\left(x_1\right) \phi\left(x_2\right) \rangle = 
	-\frac{\tilde\lambda}{8}\Delta_{F}(0) \Delta_{F}(0) \int \Delta_F(z-x_2)\Delta_F(z-x_1) d^3z.
	\eea
	Just as in \eqref{feynzero}, \textcolor{black}{here} we regulate \textcolor{black}{the divergent factor $\delta^2(0)$ appearing in $\Delta_F(0)$ by introducing the spatial lattice cutoff $a$, ie. setting $\Delta_F(0) = -\frac{i}{2m a^2}.$ Adding this to the free propagator, we get the first order corrected two-point 1-PI correlation function:}
	\begin{eqnarray} \label{tres2}
		\langle \phi\left(x_1\right) \phi\left(x_2\right) \rangle_{\tilde{\lambda}} = \frac{i}{(2\pi)^3}\int \frac{e^{-i{p.x}}}{(\omega^2-m^2+\frac{1}{8}\tilde{\lambda}\Delta_F(0)\Delta_F(0)+i\epsilon)}d^3 p.
	\end{eqnarray}
	We see that $\langle \phi\left(x_1\right) \phi\left(x_2\right) \rangle_{\tilde{\lambda}}$ possesses a pole at
	\begin{eqnarray} \label{newm2} 
		\w^2=m^2-\frac{1}{8}\tilde{\lambda}\Delta_F(0)\Delta_F(0)=m^2_R\,,
	\end{eqnarray}
		where $m_R$ is the one-loop corrected mass, matching exactly with the one in %With the point split regularization, $\Delta_{F} (0) = -\frac{i}{2m a^2}$. 
		\eqref{2ptcont}. Here we have replaced $\Delta_F (0)$ by $- \frac{i}{2ma^2}$. Hence the one-loop corrected two-point function with the renormalized mass takes the form
		\bea{} \label{tre}
		\langle \phi\left(x_1\right) \phi\left(x_2\right) \rangle_{\tilde{\lambda}} = \frac{1}{2m_R} \delta^{2}(\mbx) e^{-i m_R|t|}.
		\eea
		This expression represents the two-point correlation function of Carrollian  $\phi^6$ theory.
		%%%
		\subsubsection{Four-point function}
		
\begin{figure}[h]
			\centering
			\includegraphics[scale=0.23]{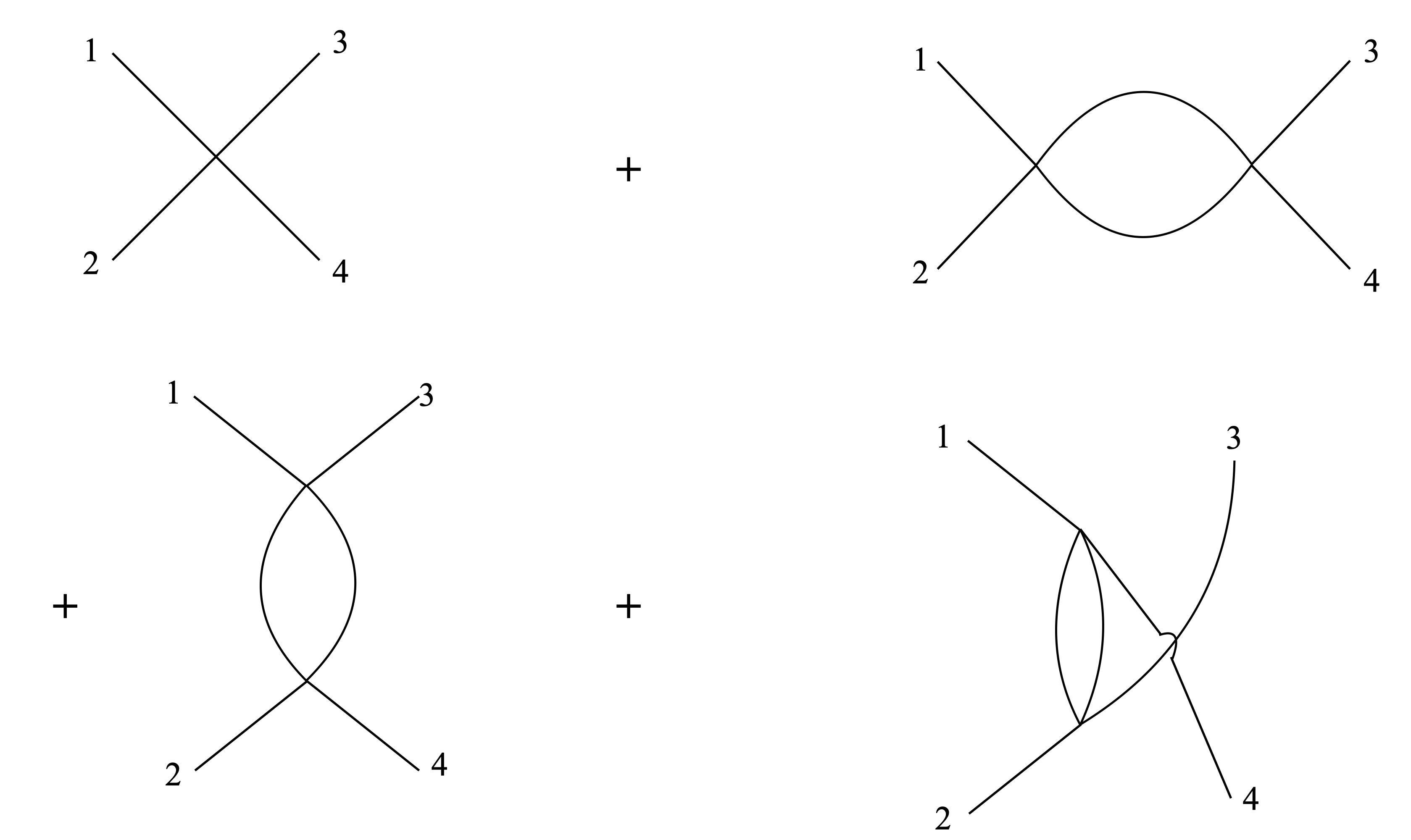}
			\caption{Feynman diagrams contributing up to two-loops in four-point function}
			\label{phi4_feyn}
		\end{figure}

		Now we focus on four-point functions in the $d=4$ theory \eqref{dfs}. Even before performing the perturbation calculation, we can comment that the higher point functions' temporal/frequency space behaviour should be exactly the same as that of a 1 dimensional harmonic oscillator with a quartic perturbation. On the other hand, the spatial part is governed by the ultralocalized contact terms. In perturbation theory, the four-point function up to second order in $\lambda$ is given by the \textcolor{black}{ connected diagrams depicted in Fig. \ref{phi4_feyn}}.
	
		It is convenient to express the correlation functions in a partial Fourier basis, where we go to frequency space by performing a Fourier transform in temporal coordinates and leave the spatial parts alone
		\begin{eqnarray} \label{fourier_part}
			\tau(\omega_1,\dots, \omega_{N}; \mathbf{x}_1,\dots, \mathbf{x}_N) = \int \dots \int dt_1 \dots dt_N e^{-i \omega_1 t_1  \ldots - i\omega_N t_N} \langle \phi(x_1) \dots \phi_N(x_N) \rangle.
		\end{eqnarray}
	\textcolor{black}{Before presenting an explicit calculation of the loop diagrams, it is worthwhile to mention using standard QFT notions applied to a 0+1 dimensional field theory, that the superficial degree of divergence of any diagram arising from the energy integrals is $=1+ E/2 -3n$. Additionally, sub-graph divergences do not occur here because they do not involve any external lines.  Here, $E$ is the number of external lines, and $n$ is the number of vertices in a particular diagram.} 
 
 The partial Fourier space version of Feynman propagator \eqref{ffh} we use to evaluate \eqref{fourier_part} is
		\begin{eqnarray}
			\tilde{\Delta}_F (\omega, \mathbf{x}) = \frac{1}{\omega^2 - m^2 +i \epsilon} \delta^3(\mathbf{x}). 
		\end{eqnarray}
		The vertex, on the other hand, contributes a factor of $-i \lambda$ as usual. For the loop diagrams, the coincident contact terms render a factor of $\delta^3(0)$ as in the case of the two-point function, and we regularize it by our splitting prescription by replacing it with $1/a^3$. That is the only source of singularity, as obviously there is no UV divergence appearing from the loop integrals. For instance in the `s-channel', i.e. the second diagram in the Fig. \ref{phi4_feyn}, the loop integral is
		\begin{eqnarray}
			\lim_{\epsilon \rightarrow 0} \int \frac{d\omega}{(\omega^2 - m^2 +i \epsilon)\left( (\omega_1 + \omega_2 -\omega)^2 - m^2 +i \epsilon \right)} = \frac{2 \pi i}{m(4m^2 - (\omega_1 + \omega_2)^2)} 
		\end{eqnarray}
		Considering the tree-level vertex and all three channels at one-loop, we summarize the four-point function in the partial Fourier space as
		\begin{eqnarray} \label{tauform}
			{\tau}\left(\{\omega_i\},\{\mathbf{x}_i\}\right) &=&  -  i\left(\prod_{i=1}^4 \frac{i}{\omega_i^2 - m^2 +i \epsilon}\right)(2\pi) \delta\left(\sum_{i=1}^{4} \omega_i\right)\delta^3(\mbx_1 - \mbx_2)\delta^3(\mbx_1 - \mbx_3)\delta^3(\mbx_1 - \mbx_4) \non \\
			&&\left( \lambda  - \frac{\lambda^2}{2 m a^3} \left( \frac{1}{s-4m^2} + \frac{1}{t-4m^2} +\frac{1}{u-4m^2}\right) \right), 
		\end{eqnarray}
		where $s= (\omega_1 + \omega_2)^2$, $t = (\omega_1 + \omega_3)^2 $ and $u =  (\omega_1 + \omega_4)^2$.
		
		To check the Ward identity for a supertranslation transformation $\delta_{f}\phi = f(\mathbf{x}) \partial_t \phi$ for this four-point function, we use the inverse partial Fourier transform
		\begin{eqnarray} \label{4pt_Ward}
			\delta_f \langle \phi(x_1)\dots\phi(x_4) \rangle = \sum_{k=1}^4 f(\mathbf{x}_k)\partial_{t_k}\left( \frac{1}{(2\pi)^4} \int  d\omega_1 \dots d\omega_4 e^{i \omega_1 t_1 \dots +\omega_4 t_4 } \tau\left(\{\omega_i\},\{\mathbf{x}_i\}\right) \right)
		\end{eqnarray}
		This identically vanishes due to the spatial delta functions, forcing all spatial insertions to coincide. The steps for this proof is given in the Appendix \ref{Proof_Ward}.
		
		We explicitly checked the above supertranslation Ward identity for the four-point function in $\phi^4$ theory \textcolor{black}{up to one-loop in perturbation theory}. However, \textcolor{black}{ as discussed in Appendix \ref{Proof_Ward}, this holds true even after including higher loop diagrams appearing in higher order in perturbation theory.} This includes the special cases for the $\phi^6$ theories we discussed earlier, in $d=3$ dimensions.
		
		The Ward identity for contributions from disconnected diagrams works pairwise for disconnected parts, as all the spatial insertions are now not forced to be coincidental.
		%%%%%
		\section{Flowing of the coupling constants}\label{sec5}
		
		Consider a class of scalar field theories in $d$ dimensions. The Carroll invariant Euclidean action takes the form
		\begin{equation} \label{action1}
			S_0[\phi]= - \int d^dx \left( \frac{1}{2} \partial{_t}\phi \partial{_t}\phi+\frac{1}{2}{m_0}^2\phi^2+\sum_{n=4}^{\infty}\lambda_{0,n}\phi^n\right).
		\end{equation}
		Here $m_0$ is the mass, and $\lambda_{0,n}$ are the coupling constants at the energy scale at which the theory \eqref{action1} is defined. In this Euclidean action, we keep only the $\mathbb{Z}_2$ symmetric terms so that the Hamiltonian is bounded from below. We briefly describe our proposed Wilsonian renormalization procedure for the Carrollian field theories described above. 
		
		One key feature in Carrollian field theories of the form (\ref{action1}) is that there are no spatial derivatives. Hence conveniently, we can expand the field in terms of energy modes
		\begin{equation} \label{enegy modes}
			\phi(t,\mathbf{x})= \int \frac{d\omega}{2\pi}e^{i\omega t}
			\phi_\omega \left(\mathbf{x}\right).
		\end{equation}
		As usual, for our theory, the renormalization group (RG) procedure will involve integrating out high-energy modes, rescaling energy, and rescaling the fields. However, it is important to note here that, unlike Lorentzian theories, the high-energy modes are not equivalent to short wavelength ones. Hence, the RG procedure we carry out here is tailor-made for Carrollian theories of type \eqref{action1} and respects Carroll symmetry at each step. This has drastic consequences in terms of scaling of operators, as we will see below. We should also point out that, due to the decoupling of spatial scale and energy scale, the RG problem becomes closely analogous to the one in a 0+1 dimensional QFT, i.e. in a quantum mechanical anharmonic oscillator system. Hence, the following discussion can be compared to the Callan-Symanzik flow equations for $O(N)$ symmetric anharmonic oscillator models presented recently in \cite{Giorgini:2020acn}. We would also like to point out that renormalization group flow for systems with a single degree of freedom was studied in \cite{Polonyi:1994pn}, where the couplings ran with a time-scale (equivalent to an energy cutoff scale).\\

        The path integral for the free theory is of the form
        \begin{eqnarray}
            Z_{\mathrm{free}} = \int \mathcal{D}\phi\, e^{-\int d^{d-1} \mbx \,d\omega \,\frac{1}{4 \pi}\left( \omega^2  + m^2\right)\phi_{\omega}(\mbx)\phi_{-\omega}(\mbx)}, \nonumber
        \end{eqnarray}
But, as is well known \cite{zinn1993quantum}, this diverges, even for the $0+1$ dimensional theory, unless one puts an upper cutoff in the $\omega$  variable. Hence we define our theory with an upper cutoff in energy $\Omega$ i.e. $\phi_\omega =0  , |\omega| > \Omega$. We also introduce an intermediate cutoff $\Omega'= \frac{\Omega}{\zeta}$, where $\zeta>1$ and split the energy modes as low-energy ($\phi^-_{\omega}$) and high-energy ones ($\phi^+_{\omega}$)
		\textcolor{black}{\begin{equation} \label{eqn four}
			\phi_\omega =  
            \begin{cases}
			    \phi_\omega ^-, & \mbox{ if } \omega < \Omega',\\ \phi_\omega^+, & \mbox{ if } \Omega' <\omega < \Omega.
			\end{cases}
		\end{equation}}
		Accordingly, the Euclidean action can be decomposed into high-energy and low-energy modes
		\begin{equation} \label{eqn five}
			S[\phi_{\omega}]= S_0[\phi_\omega^-]+S_0[\phi_\omega^+]+S_I[\phi_\omega^-,\phi_\omega^+].
		\end{equation} 
		Here $S_I[\phi_\omega^-,\phi_\omega^+]$ involves terms that mix the high and low-energy modes. After integrating out the high-energy modes, the partition function becomes
		\begin{equation} \label{eqn eight}
			Z= \int \mathcal{D}\phi^- ~e^{-S'[\phi^-]},
		\end{equation}
		where the Wilsonian effective action $S'[\phi^-]$ is given by
		\bea{}&& \label{eqn nine}
		e^{-S'[\phi^-]}=e^{-S_{0}[\phi^-]} \int \mathcal{D}\phi^+ 
		e^{-S_0[\phi^+]-S_I[\phi^-,\phi^+]}.
		\eea
		At the next step of the Wilsonian procedure, we rescale the intermediate energy scale $\omega'=\zeta \omega$. To obtain the scaling $w$ of the fields with energy scale: $\left( \phi^-_\omega \left( \mathbf{x} \right) = \zeta^w \phi'_{\omega'} \left( \mathbf{x} \right) \right)$ refer to the free Euclidean theory:
		\begin{eqnarray} \label{just}
			S_0[\phi] = \int d^dx \left(\frac{1}{2} \dot{\phi}^2 + \frac{1}{2} m_0^2 \phi^2 \right).
		\end{eqnarray}
		After performing the trivial gaussian path integral and the rescaling, one finds that the the kinetic term remains canonical only if $w = 3/2$. Hence the fields scale with energy as: $\phi^-(t,\mathbf{x}) = \zeta^{1/2} \phi'(t',\mathbf{x})$ or $\phi^-_\omega (\mathbf{x}) = \zeta^{3/2} \phi'_{\omega'} (\mathbf{x})$. $\mathbf{x}$ behave as spectator coordinates in this scaling process. This scaling behaviour \textcolor{black}{is the same as that of RG studies in the context of quantum mechanics \cite{Polonyi:1994pn, Giorgini:2020acn}}. \textcolor{black}{In the present context it} is intrinsic to understanding flow in energy scale in Carrollian \textcolor{black}{theories.} \textcolor{black}{Hence} the {\it Carroll dimensions} of fields are different than the canonical dimensions, defined by the infinitesimal transformation $t\partial_t +\mathbf{x}\cdot \nabla$, which essentially are $d$ dependent. Mass, on the other hand, scales as expected: $m^2(\zeta) = \zeta^2 m^2_0$. The tree-level scaling of the operators $\lambda_{0,n} \phi^n$ reflects in the flow $\lambda(\zeta) = \zeta^{1+n/2} \lambda_{0,n}$ \footnote{To be contrasted with Lorentz covariant scaling: $\lambda(\zeta) = \zeta^{d-n(d-2)/2} \lambda_{0,n}$ in $d$ space-time dimensions.}. Hence all of these couplings are relevant in the Carrollian sense at tree-level.
		
		A curious upshot of this scaling behaviour of the fields is the emergence of Lorentz invariance. To understand this, let's add to \eqref{just} a deformation $c^2 \nabla \phi \cdot \nabla \phi $, for $c^2 \ll 1$. As a consequence of the above scaling law, $c^2$ scales as: $c^2 \rightarrow \zeta^2 c^2$. This implies \textit{that as one goes deeper in to the IR, the $c^2 \nabla \phi \cdot \nabla \phi $ deformation is relevant in Carroll sense and at a scale well in the IR, one can reach $c =1$ with Lorentz symmetry emerging.} This is similar to the $z=2$ (space-time anisotropic scaling exponent) Lifshitz scalar \cite{Horava:2009uw} flowing to $z=1$ in the IR making Lorentz symmetry emergent \cite{Chadha:1982qq}. \textcolor{black}{In fact any operator of the form $\phi \nabla^{2\gamma} \phi$, for $\gamma >1$ is relevant. Hence, such a Carroll breaking deformation can lead one to Lifshitz fixed points for different values of $\gamma$.}
  
  In addition to that, this observation reiterates the newer findings \cite{Bagchi:2023ysc} in the context of fluids at very high-energy scale. Fluids at that scale are described by Bjorken flow moving ultra-relativistically and has emergent Carrollian symmetry. Similar descriptions of Carroll fluids on non-flat null manifolds have been studied in the context of flat holography in \cite{Ciambelli:2018wre}.
		%As a result, from the definition of the frequency modes, we get the scaling of the fields: $\phi'_{\omega'}=\zeta^{-\frac{d+2}{2}}\phi_{\omega}^-$ such that both $S'$ and $S$ have the same cut off and the kinetic term keeps the canonical normalization.
		%We now do a perturbative expansion as,
		%\begin{equation}
		%      S'[\phi^-]= S_0[\phi^-]+\langle S_{I} \rangle_{} -\frac{1}{2} [\langle S_{I}^2 \rangle_{} -\langle S_{I}\rangle ^2_{} ].
		% \end{equation}
		%Here $S_0[\phi^-]$ has $\Omega$ as the cutoff where $S'[\phi]$ has $\Omega'$ as its cutoff.
		%Thus we rescale $\omega'=\zeta \omega$ so that both $ S_0$ and $S_I$ have the same cutoff $\Omega$; We also rescale the field as $\phi_{\omega}'=\zeta^{\frac{-3}{2}}\phi_{\omega}^-$ such that both $S_0$ and $S_I$ has the same functional form which we need to maintain during a Renormalisation Group Flow procedure.
		
		%Thus  Renormalization Group procedure  includes integrating out high-energy modes, rescaling energy, and rescaling the fields. The crux is to understand how the mass and the coupling constant get scaled under RG flow. We then carry out first and second-order calculations.
		%At the first order, we calculate $\langle S_{I} \rangle_{}$. %We then expand $\phi^d$ in Fourier modes. %We  take the action and add a source term to it. Then we differentiate it wrt source two terms and we get  the relation \begin{equation} \label{eqn sixteen}.
		While integrating out higher energy modes perturbatively, we need the following two-point function at coincident spatial points ($\mathbf{x} \rightarrow \mathbf{x'}$)
		% Analogous to Lorentzian field theories, we make use of the following relation to find out the first order and second order corrections.
		\begin{equation} \label{eqn sixteen}
			\langle \phi_\omega^+ \left(\mathbf{x}\right)\phi_{\omega'}^+\left(\mathbf{x'}\right)\rangle_+ :=
			\int \mathcal{D} \phi^+ \phi_\omega^+ \left(\mathbf{x}\right)\phi_{\omega'}^+\left(\mathbf{x'}\right) e^{-S_0[\phi^+]}= 2\pi \delta(\omega+\omega') G_0(\omega) \delta^{d-1}\left(\mathbf{x}-\mathbf{x'}\right),
		\end{equation}
		where $G_0(\omega) = \frac{1}{\omega^2 + m_0^2}$ is the Green's function of the free theory. In order to make sense of the divergence coming from the $\delta^{d-1}$ function, we use the same lattice regularization as used in Section \ref{lattice} to infer
		\begin{eqnarray}\label{eqn sixteen prime}
			\lim_{\mathbf{x} \rightarrow \mathbf{x'}} \langle \phi_\omega^+ \left(\mathbf{x}\right)\phi_{\omega'}^+\left(\mathbf{x'}\right)\rangle_+ = 2\pi \delta\left(\omega+\omega'\right) G_0\left(\omega\right) \frac{1}{a^{d-1}},
		\end{eqnarray}
		where $a$ is the lattice scale.
		%However unlike the Lorentzian theories the two-point functions are now evaluated at the same space points and hence  (\ref{eqn sixteen}) gives us a divergent quantity. So we do a point splitting by taking ${|x-x'|} = a$, where $a$ is the lattice spacing as described in the previous sections. Thus we modify (\ref{eqn sixteen}) as
		%\begin{equation} \label{eqn sixteen prime}
		%    \langle \phi_\omega^+ \left(\mathbf{x}\right)\phi_{\omega'}^+\left(\mathbf{x}\right)\rangle_+ = 2\pi \delta\left(\omega+\omega'\right) G_0\left(\omega\right) \frac{1}{a^{d-1}},
		%\end{equation}
		%In Lorentz invariant theory in d dimension, \ref{eqn sixteen} is modified as
		%\begin{equation}
		% \langle \phi_k^+ \phi_{k'}^+\rangle_+ = (2\pi)^d \delta^d(k+k') G_0(k)  
		%\end{equation}
		%where $G_0(K)$ is the Greens function.
		%Then we rescale energy and fields to complete the RG procedure.%
		%At the second order we evaluate$-\frac{1}{2} [\langle S_{I}^2 \rangle_{} -\langle S_{I}\rangle ^2_{} ]$. We also use Wick's theorem to perform the second-order calculations.
		Finally, comparing the Wilsonian Effective Free Theory with the starting action, we can read off the change in the mass and the coupling constants order by order. These are expressed in terms of the beta functions.
		%Finally, we will find beta functions which are first-order differential equations used to understand the flow of coupling constants given by
		\begin{equation}
			\beta = -\Omega' \frac{dg}{d\Omega'}=-\frac{dg}{d s} \hspace{3em} \mbox{where } s= \ln(\Omega/\Omega') = \ln \zeta
		\end{equation}
		To understand how this specific RG flow in Carrollian theories compares with Lorentzian ones, we will focus on marginal local couplings in Lorentzian theories. A couple of such examples are $\phi^4$ coupling in $d=4$ and $\phi^6$ in $d=3$. We will explicitly calculate the beta functions in these theories up to first order in perturbation theory and search for nontrivial fixed points in the space of parameters.
		% We will explicitly calculate the beta functions for a couple of theories below in the \textit{ultraviolet} (i.e. small $s$) limit perturbatively and analyze the flows.
		%where $\Omega'=\frac{\Omega}{\zeta}=\Omega e^{-s}$. 
		
		%As an example, we consider $\phi^4 $ theory in $d=4$ dimension and $\phi^4+\phi^6$ in $d=3$ dimension.
		
		\subsection{$\phi^4$ theory in $d=4 $ dimension} \label{sec:5.1}
		
		Consider the Euclidean form of the action (\ref{dfs}) in $d=4$ dimensions
		\begin{equation} \label{eqn ten}
			S_0[\phi]=\int d^4x \left[\frac{1}{2} \partial{_t}\phi \partial{_t}\phi+\frac{1}{2}{m_0}^2\phi^2+\dfrac{\lambda_{0}}{4!}\phi^4\right],
		\end{equation}
		where $m_0$ is the mass and $\lambda_0$ is the coupling constant for $\phi^4$ theory. From the discussion above, we now observe that both the terms ${m_0}^2\phi^2$ and $\lambda_{0}\phi^4$ are relevant at tree-level. Hence any deformation in the $\{m, \lambda\}$ coupling sub-space, near the Gaussian fixed point $m = 0 = \lambda$ makes it flow away from the point with RG.
		Carrying out the Wilsonian RG process described above, we see that at \textcolor{black}{linear} order \textcolor{black}{in} $\lambda_0$, the mass flow is given by
		\begin{equation} \label{eqn seventeen}
			m^2(\zeta)=\zeta^2\left[m_0^2+\frac{\lambda_0}{2\pi}\frac{1}{a^3}\int_{\frac{\Omega}{\zeta}}^{\Omega}
			\frac{d\omega}{\omega^2+m_0^2}\right],
		\end{equation}
		whereas the effective lower energy coupling is   
		\begin{equation} \label{eqn ninteen}
			\lambda(\zeta)= \zeta^3 \left[\lambda_0-\frac{3\lambda_0^2 }{2\pi} \frac{1}{a^3} \int_{\frac{\Omega}{\zeta}}^{\Omega}\frac{d\omega}{(\omega^2+m_0^2)^2}\right]. 
		\end{equation}   
		The integrals appearing above are elementary. Expanding the above loop integrals in a power series around $\zeta=1$, the beta functions turn out to be:
		\begin{eqnarray} \label{beta_m}
			\beta_m = - \frac{dm^2}{ds} =  -2m^2-\dfrac{\lambda}{2\pi}\dfrac{1}{a^3}\dfrac{\Omega}{\Omega^2+m^2}, ~~    \beta_{\lambda}=-\frac{d\lambda}{ds}=-3\lambda+\frac{3\lambda^2}{2\pi}\frac{1}{a^3}\frac{\Omega}{(\Omega^2+m^2)^2}.
		\end{eqnarray}      
		% \begin{figure}[t]
		%\centering
		%\begin{subfigure}{0.47\textwidth}
		%\centering
		% \includegraphics{4pt_tree_quartic.pdf}
		%\end{subfigure}
		%+
		% \begin{subfigure}{0.47\textwidth}
		% \centering
		%\includegraphics{4pt_loop_quartic.pdf}
		%\end{subfigure}
		%\caption{four-point correlation function of the $\phi^4$ theory at one-loop}
		%\label{fig 2}
		%\end{figure}
		%
		%%%
		One striking feature of this result \eqref{beta_m} is that at the Gaussian fixed point ($m^2 = 0 = \lambda$),  both mass and the coupling are relevant at tree-level. This is in contrast to the Lorentzian counterpart, where the coupling is marginal at tree-level and becomes marginally irrelevant as quantum effects are included. We will explore more about the flows and fixed points in the later section.
		
		As per the Carrollian RG program described above, we consider flows only from higher energy to lower energies, and do not consider any coarse graining along the spatial dimensions. Hence, the spatial lattice parameter $a$ will be kept fixed while analyzing \eqref{beta_m}. %The integrals appearing in the one-loop correction are elementary and can be evaluated and expanded in a power series around $\zeta=1$, to extract trivially giving 0 as $\Omega \rightarrow \infty$ for any finite $\zeta$. In order to extract true flow information as we go to the infra-red, we evaluate the integrals and expand in powers of $1/\Omega$, for $\zeta - 1 \ll 1$. This analysis renders the following form of the beta functions:
		%\begin{eqnarray}\label{betaall4}
		%    \zeta \frac{d m^2}{d\zeta} = -2 m^2- \frac{\lambda}{2\pi}  \frac{1}{\Omega a^3}, ~~~~
		%    \zeta \frac{d \lambda}{d\zeta} = -\dfrac{ \lambda}{8} + \frac{1}{16\pi} \frac{\lambda^2}{(a\Omega)^3}
		%\end{eqnarray}
		Clearly, apart from the Gaussian fixed point, $m^2 = 0,  \lambda =0$ an additional nontrivial fixed point is there
		\begin{eqnarray} \label{fp1}
			m^2_{\star} =  -\Omega^2/3, \lambda_{\star} = \frac{8\pi}{9} (a\Omega)^3.
		\end{eqnarray}
        Since $a$ is independent of $\Omega$, we can choose it so that that $\lambda_{\star}$ is small and this fixed point is perturbatively accessible. To understand the flow of mass and the coupling constant near this fixed point, we perform a linearized variation of the beta functions \eqref{beta_m} around this point \begin{equation}
			m^2= m_{\star}^2 +\delta m^2 ,~~~~
			\lambda = \lambda_{\star} +\delta \lambda,
		\end{equation}
		to arrive at
		%\begin{equation}
		% \zeta \frac{d \delta m^2}{d\zeta}  = -2\delta m^2-\frac{1}{2\pi \Omega a^3} \delta \lambda, ~~~\zeta \frac{d \delta \lambda}{d\zeta} = \frac{1}{8}\delta \lambda .
		%\end{equation} 
		%Thus we get 
		\begin{equation} \label{matrix}
			-\zeta \dfrac{d}{d\zeta} \begin{pmatrix} \delta m^2\\ 
				\delta \lambda\\\end{pmatrix} = \begin{pmatrix} -1 && -\dfrac{3}{4\pi \Omega a^3}\\ -8\pi \Omega a^3 && 3\\ \end{pmatrix}
			\begin{pmatrix} \delta m^2\\ \delta \lambda\\ \end{pmatrix}.
		\end{equation}
		The eigenvalues of the above matrix are $1 \pm \sqrt{10}$. Hence one of the directions is marginally relevant and the other is marginally irrelevant and the fixed point is not stable. 
		%Thus we evaluated the fixed point analysis in $d=4$ dimension.
		
		\subsection{$\phi^4+\phi^6$ theory in $d=3$ dimension} \label{sec:5.2}
		
		We start off with \textcolor{black}{an} action of the form
		\begin{equation} \label{eqn twenty one}
			S_0[\phi]=\int d^3x \left[\frac{1}{2} \partial{_t}\phi \partial{_t}\phi+\frac{1}{2}{m_0}^2\phi^2+\dfrac{{\lambda_0}}{4!}\phi^4+\dfrac{\tilde{\lambda}_{0}}{6!}\phi^6\right],
		\end{equation}
		where $m_0$ is the mass, $\lambda_{0}$ is the coupling for the field $\phi^4$  and $\tilde{\lambda}_{0}$ is the coupling constant of the field $\phi^6$. Here, unlike (\ref{dfs3}), we keep all the relevant couplings in $d=3$. 
		%The Feynman diagrams for the first order and second are given  by Fig. \ref{fig 3} and \ref{fig 4}.
		%Now using the same procedure as we have done for the previous section, we calculate mass correction and correction to the couplings for $\phi_6$ theory in $d=3$ dimension.
		At \textcolor{black}{ $\mathcal{\lambda_0}$} , the effective mass is given by
		\bea{}\label{eqn twenty two}
		m^2(\zeta)=\zeta^2\left[m_0^2+\frac{{\lambda_0}}{2\pi}\frac{1}{a^2}\int_{\frac{\Omega}{\zeta}}^{\Omega}
		\frac{d\omega}{\omega^2+m_0^2} +\frac{ \tilde{\lambda}_0}  {16\pi^2 } \frac{1}{a^4} \left(\int_\frac{\Omega}{\zeta}^{\Omega}\frac{d\omega}{(\omega^2+m_0^2)}\right)^2\right],
		\eea
		and the corresponding beta function is 
		\begin{equation} \label{eqn twenty three}
			\beta_m =-\frac{dm^2}{ds} =-2m^2-\frac{\lambda}{2\pi}\frac{1}{a^2} \frac{\Omega}{\Omega^2+m^2} - \frac{ \tilde{\lambda}}{16\pi^2} \frac{1}{a^4} (2s)\frac{\Omega^2}{(\Omega^2+m^2)^2}.
		\end{equation}
		We calculate the beta function for each coupling separately using the same RG procedure. The low-energy effective $\lambda$ and the corresponding beta function are given by:
		\begin{equation} \label{eqn twenty five}
			{\lambda}= \zeta^3\left[\lambda_0 -\frac{3{\lambda_0}^2}{2\pi} \frac{1}{a^2} \int_{\frac{\Omega}{\zeta}}^{\Omega} \frac{d\omega_4}{(\omega_4^2+m_0^2)^2} \right], ~~ \beta_{\lambda} =-\frac{d{\lambda}}{ds} = -3\lambda +\frac{ 3\lambda^2 }{2\pi}\frac{1}{a^2} \frac{\Omega}{(\Omega^2+m^2)^2}.
		\end{equation}
		%and the corresponding beta function is given by
		%\begin{equation} \label{eqn twenty five}
		%  \beta_{\lambda} =-\frac{d{\lambda}}{ds} = -\dfrac{ {\lambda}}{8}+\frac{ {\lambda}^2 }{(16\pi)}\frac{1}{a^2} \frac{\Omega}{(\Omega^2+m^2)^2}.
		%\end{equation}
		Similarly, the effective low-energy $\tilde\lambda$ and the corresponding beta function are
		\begin{eqnarray} \label{eqn twenty seven}
			&&  \tilde{\lambda} =  \zeta^4 \left[\tilde{\lambda}_0 -\frac{15\tilde{\lambda}_0^2}{16\pi^2} \frac{1}{a^4} \int_{\frac{\Omega}{\zeta}}^{\Omega}
			\frac{d\omega'_1 d\omega'_2}{({\omega'}_1^2 + m_0^2)^2 ({\omega'}_2^2 + m_0^2)}\right], \non \\ && \beta_{\tilde\lambda} = -4\tilde{\lambda} + \frac{15\tilde{\lambda}^2\,s}{8\pi^2}  \frac{1}{a^4}\frac{\Omega^2}{(\Omega^2+m^2)^3} .
		\end{eqnarray}
		To search for fixed points in the parameter space, we solve the beta functions \eqref{eqn twenty three},\eqref{eqn twenty five} and \eqref{eqn twenty seven} set \textcolor{black}{to} zero for negligibly small $s$. Apart from the Gaussian fixed point, there is only one other solution which is perturbatively accessible
		\begin{eqnarray} \label{fp2}
			m^2_{\star} = -\Omega^2/3, ~~ a\lambda_{\star} = \frac{8\pi}{9} (a \Omega)^3, ~~ \tilde{\lambda}_{\star} = 0.
		\end{eqnarray}
		The flow in the neighbourhood of this fixed point is given by
		\begin{equation} \label{matrix2}
			-\zeta \dfrac{d}{d\zeta} \begin{pmatrix} \delta m^2\\ 
				\delta \lambda\\
				\delta\tilde\lambda\end{pmatrix} = \begin{pmatrix} -1& &-\frac{3}{4\pi \Omega a^2}& & 0\\ -8\pi a^2 \Omega & &3 & &0\\ 0& &0& &-4\end{pmatrix}
			\begin{pmatrix} \delta m^2\\ \delta \lambda\\ \delta\tilde\lambda\end{pmatrix}.
		\end{equation}
		The eigenvalues of the above matrix are $1\pm \sqrt{10}$ in the $\{ \delta m^2 , \delta \lambda \}$ subspace and $-4$ in the $\delta \tilde{\lambda}$ space. Hence the $\delta \tilde{\lambda}$ direction is relevant, whereas for the $\delta \tilde{\lambda} =$ constant submanifold has one relevant and one irrelevant direction. Therefore, this fixed point is unstable.
		%and the corresponding beta function is given by
		%\begin{equation} \label{eqn twenty seven}
		% \beta_{\tilde\lambda} = -\dfrac{\tilde{\lambda}}{180}+ \frac{\tilde{\lambda}^2}{768\pi^2}  \frac{1}{a^4}(2s)\frac{\Omega^2}{(\Omega^2+m^2)^3} .
		%\end{equation} 
		At this point, a few \textcolor{black}{comments } are in order. First, the effects of $\tilde{\phi}^6$ coupling in the flow of mass \eqref{eqn twenty two} and to $\tilde{\lambda}$ itself appear at two-loop level. Hence the effects of these are small as we integrate out modes in the range $[\Omega/\zeta, \Omega]$. Second, note that the flow $\tilde\lambda$ is not affected by $\lambda$ as in Lorentzian theory. However, just as in the 4 dimensional case, the $\phi^6$ coupling is not marginal at tree-level in our Carrollian scheme of Wilsonian RG. A comparison of the beta functions of the Carrollian theories and their Lorentzian counterparts is given in Appendix \ref{appendix1}.
		
		A few takeaways from the fixed point analysis are as follows.
		\begin{itemize}
			\item \textcolor{black}{We have adopted an RG scheme suited for the Carrollian scalar theories, where energy scales are decoupled from spatial scales. This drastically differs from the Wilsonian RG for (Euclideanized) Lorentz covariant theories. As a result, operators' relevance, irrelevance or marginality in a Carrollian theory are not the same as those in a Lorentz covariant theory. For example, a $\phi^4$ coupling in a 4 dimensional Lorentzian theory is marginal but is relevant in a Carrollian theory, as is evident from \eqref{eqn ninteen}.}
			\item There exist nontrivial fixed points in Carrollian scalar field theories. One may compare the situation with the Wilson-Fisher fixed point for scalar field theories, encountered in the `$\varepsilon$' expansion of dimensional regularization. \textcolor{black}{First, the mass term appearing has a negative sign, and Second, one of the flow directions is marginally relevant while the other is marginally irrelevant. Both these features are prominent in the Wilson-Fisher fixed point of the relativistic $\phi^4$ theory.}
   
			\item The interplay between the RG scheme and Carrollian symmetry manifestly keeps the supertranslation symmetry intact at each point of the flow.
			%\item However, turning on a small Carroll violating deformation at the Gaussian fixed point takes it away and well into the IR, Lorentz invariance emerges.
		\end{itemize}
		
		\section{Conclusions}\label{conclusions}
		
		In the present article we probed into the quantization of a simple interacting Carroll scalar with manifest ultralocal behaviour at one-loop level. The study was inspired by ultralocal features of celestial CFTs. The supertranslation invariance, although intuitively plausible, was explicitly shown to hold even after incorporation of one-loop quantum effects. A number of results we presented in the present paper including the existence of a disentangled ground state, the structure of the one-loop two-point function, and supertranslation Ward identities for $n$-point correlators can be argued to hold true without digging deeper into the technical details. This is because we did not have any spatial gradient term in the theory. In Carrollian physics literature \cite{Henneaux:2021yzg}, these manifestly ultralocal theories are named as the ``electric" type. However, there are examples of Carrollian gauge theories \cite{Duval:2014uoa, Bagchi:2016bcd, Islam:2023rnc}, fermions \cite{Bagchi:2022eui}, and gravity \cite{Campoleoni:2022ebj} containing spatial derivatives, named as the ``magnetic" type. It would be crucial to check which of the results presented in this paper are universal for Carrollian field theories; i.e. if they continue to hold for theories of magnetic type.
		
		In particular, there exists lattice models in two spatial dimensions with nearest-neighbour couplings (hence not ultralocal) which show dispersionless and hence Carrollian features \cite{Bagchi:2022eui}. From a condensed matter/ quantum information perspective it is crucial to see the entanglement structure of this type of Carrollian many-body theories. That would be a concrete probe into the power of Carrollian symmetries in higher dimensions (the $\left(1+1\right)$-dimensional case at critical point and its universality of entanglement has already been established in \cite{Bagchi:2014iea} using symmetry principles alone).
		
		The issue of the conformal Ward identity at the level of the tree-level time-ordered two-point function is subtle, since the $m \rightarrow 0$ limit is not trivial. Rather one has to extract the physically meaningful piece at the massless limit to see conformal invariance. Similar nontrivialities plague the higher point functions even at the tree-level. It would be crucial to check the global conformal Ward identities for higher point functions. Especially for the $d=3$ theory, apart from the global conformal symmetry the extended BMS group demands symmetry with respect to the 2D stress tensor, which includes the super-rotation generators as well. From the perspective of Carrollian holography proposal, the study of super-rotation Ward identities and their anomaly structure at loop level would be an important component of further studies.
		
		We invested a considerable amount of effort in understanding the role of the energy scale for the ``electric"-type Carrollian scalar. Contrary to the standard Wilsonian point of view of scales, we worked with the idea that at least for the present set of theories, the concept of spatial scale is decoupled from that of the energy scales. Hence it becomes practical to study the scaling properties of operators from the point of view of just energy scales. As a direct consequence of this approach, these properties become essentially different from the canonical scaling dimensions. Pursuing further along these lines, we focused on the IR flow tailor made for electric-type Carrollian theories. We observed that a deformation like $c^2 \nabla\phi \cdot \nabla \phi$ near the Gaussian fixed point is relevant, and sufficient amount of (finite value of the flow parameter) flow towards the IR takes one to $c=1$, thus instating Lorentz invariance. As a complementary point of view, it would be interesting to see Carroll symmetry emerging, starting from a Lorentz invariant theory with a marginal deformation, if possibly one can integrate out lower energy modes. Such a complementary viewpoint is going to appear soon in \cite{Aritra_upcoming}. Notably, a concept of dialing a current-current deformation strength in a CFT, to flow from Lorentzian to the Carrollian regime was alluded to in \cite{Bagchi:2022nvj}.
		
		Up to lowest order in perturbative analysis, we scanned for RG fixed points in the parameter spaces, albeit restricted to a substantially small dimensional subspace. The motivation behind choosing $\phi^4$ coupling in $d=4$ and $\phi^4 + \phi^6$ in $d=3$ was to keep only those operators which are relevant/marginal in the Wilsonian sense. However, these notions no longer hold true in the RG program we worked with. Even within this small set of parameters we unveiled new fixed points apart from the Gaussian ones. These are strictly different from the Wilson-Fisher ones, which are close to the Gaussian ones. A number of future studies can be proposed from here: i) It would be interesting to study sigma models with Lie group internal symmetries to uncover richer features of spontaneous symmetry breaking at nontrivial fixed points like this. (ii) These fixed points are not conformal critical points from the traditional sense of criticality in Lorentzian (or their Euclidean versions) sense. However, now letting the spatial scale $a$ vary, one should investigate Ward identities for dilatation $t \partial_t + \mathbf{r} \cdot \nabla$ at these points. (iii) Since none of the power-law ultralocal operators respecting Carroll invariance is irrelevant or marginal and since, from a practical point of view, the loop integrals are much more tractable, it would be plausible to enlarge the parameter space by including more relevant operators and then search systematically for more nontrivial fixed points.
		\section*{Acknowledgements}
		It is a pleasure to thank Arjun Bagchi, Aritra Banerjee and Shankhadeep Chakrabortty for illuminating discussions and inputs regarding the draft. The work of RB is supported by the following grants from the Science and Engineering Research Board (SERB), India: CRG/2020/002035, SRG/2020/001037 and MTR/2022/000795. KB would like to thank CRG/2020/002035 for support. AR is supported by an appointment to the Junior Research Groups (JRG) Program at the Asia Pacific Centre for Theoretical Physics (APCTP) through the Science and Technology 
		Promotion Fund and Lottery Fund of the Korean Government and  by the National Research Foundation of Korea (NRF) grant funded by the Korea government (MSIT) (No. 2021R1F1A1048531). AM would like to thank the support of the Department of Physics at BITS-Pilani, K. K. Birla Goa campus through the grant SERB CRG/2020/002035, and the Royal Society University Research Fellowship of Jelle Hartong through the Enhanced Research Expenses 2021 Award grant number: RF/ ERE/210139 where part of this work was conducted.
		
		\appendix 
		\section{Derivation of \eqref{correlator_discrete_phi6}} \label{Proof_2pt}
		For a hexic perturbation $\tilde{\Lambda} X^6/6!$ to a harmonic oscillator the $n$th perturbed energy and perturbed state vector up to first order in perturbation are given by
		\begin{eqnarray}
			\tilde{E}_n = E_0 + \frac{\tilde{\Lambda}}{6!} \langle n| X^6 |n \rangle  , ~~~|\tilde{n}\rangle = |{n}\rangle + \frac{\tilde{\Lambda}}{6!}\sum_{m \neq n} \frac{\langle m|X^6 |n\rangle }{E_n -E_m} |m\rangle . 
		\end{eqnarray}
		Since the operator $X$ is a sum of creation and annihilation operator, it's easy to see that the matrix elements $\langle m|X^6 |n\rangle $ appearing in the state perturbation are non-zero when the positive integer $m \in \mathcal{C}_n = \{m\Big| |m-n| \leq 6; m\neq n  \}$. Keeping in mind that we need $|\langle \tilde{0}| X | \tilde{n} \rangle|^2 $ up to first order in $\tilde\Lambda$ following simplified form is useful
		\begin{eqnarray} \label{statesnew}
			%  |\tilde{0} \rangle = |0\rangle + \alpha_1\tilde{\Lambda} |6 \rangle, ~~ 
			|\tilde{n}\rangle = |{n}\rangle + \tilde{\Lambda}\sum_{m \in \mathcal{C}_n} \beta_m |m\rangle ,
		\end{eqnarray}
		where $\alpha_1, \beta_m$ are coefficients involving matrix elements $\langle m|X^6 |n\rangle$. A quick look at \eqref{statesnew} now leads us to conclude the following $\tilde{\Lambda}$ dependencies
		\begin{eqnarray}
			\langle \tilde{0} | X | \tilde{1} \rangle = \mathcal{O}(\tilde{\Lambda}^0) + \mathcal{O}(\tilde{\Lambda}^1), ~~  \langle \tilde{0} | X | \tilde{n} \rangle \Big|_{n \neq 1} = \mathcal{O}(\tilde{\Lambda}^1).
		\end{eqnarray}
		Hence out of all elements $\langle \tilde{0} | X | \tilde{n} \rangle$, only the one for $n=1$ contributes a linear $\tilde{\Lambda}$ term in $|\langle \tilde{0} | X | \tilde{n} \rangle|^2$.
		
		For easiest computation of the elements like $\langle m|X^6 |n\rangle$, we used {\verb|Mathematica|} \cite{Mathematica} to compute integrals of the form:
		$$ \sim \int e^{-x^2} x^6 H_m(x) H_n(x).$$
		Following these steps we end up with the final expression:
		$|\langle \tilde{0} | X | \tilde{1} \rangle|^2 = \frac{\hbar}{2 M \omega}\left(1 - \frac{\tilde\Lambda}{64} \frac{\hbar^2}{M^3\omega^4}\right)$. \textcolor{black}{ Higher order perturbative corrections are of the form of $\mathcal{O}\left(\frac{\tilde{\Lambda}^2 \hbar^4}{M^6 \omega^6}\right)$.}
        %%%
		\section{Proof of vanishing of equation \eqref{4pt_Ward}}\label{Proof_Ward}
  
		The right hand side of \eqref{4pt_Ward} is
		\begin{eqnarray} \label{WardRHS}
			\sim   \int  d\omega_1 \dots d\omega_4 e^{i \omega_1 t_1 \dots +\omega_4 t_4 } \left(\sum_{k=1}^4 f(\mathbf{x}_k)i\omega_k \right)\tau\left(\{\omega_i\},\{\mathbf{x}_i\}\right), 
		\end{eqnarray}
		But the form of $\tau$, according to \eqref{tauform} is
		\begin{eqnarray} \label{tauform}
			{\tau}\left(\{\omega_i\},\{\mathbf{x}_i\}\right) = g(\{\omega_i\}, \lambda, m, a)\delta\left(\sum_{i=1}^{4} \omega_i\right)\delta^3(\mbx_1 - \mbx_2)\delta^3(\mbx_1 - \mbx_3)\delta^3(\mbx_1 - \mbx_4), 
		\end{eqnarray}
		where $g$ is a meromorphic function of the $\omega$-s, which has isolated poles. Now the three spatial delta functions force the points $\mathbf{x}_2, \mathbf{x}_3, \mathbf{x}_4$ to be identified with $\mathbf{x}_1$. Hence \eqref{WardRHS} becomes
		\begin{eqnarray} \label{vanish}
			f(\mathbf{x}_1) && \int  d\omega_1 \dots d\omega_4 e^{i \omega_1 t_1 \dots +\omega_4 t_4 } \left(\sum_{k=1}^4 i\omega_k \right)\delta\left(\sum_{l=1}^{4} \omega_l\right) \non \\ && g(\{\omega_i\}, \lambda, m, a)\delta^3(\mbx_1 - \mbx_2)\delta^3(\mbx_1 - \mbx_3)\delta^3(\mbx_1 - \mbx_4) =0
		\end{eqnarray}
 \textcolor{black}{The above arguments can be generalized to $N$-point correlators in a Carrollian scalar theory with the interaction of the form $\sim \lambda \phi^m$. Let now, $\tau_N \left(\{\omega_i\},\{\mathbf{x}_i\}\right)$ as in \eqref{fourier_part}, be the correlation function correct to any finite order in perturbation theory, consisting only of connected diagrams with 1 or more loops involved. If we assume that none of the loop integrals in $\omega$ diverge, then the structure of $\tau_N$ takes the form similar to 
 \eqref{tauform}
		\begin{eqnarray} \label{tauform_N}
			{\tau_N}\left(\{\omega_i\},\{\mathbf{x}_i\}\right) = g_N(\{\omega_i\}, \lambda, m, a)\delta\left(\sum_{i=1}^{N} \omega_i\right)\prod_{i=2}^{N} \delta^3(\mbx_1 - \mbx_i). 
		\end{eqnarray}
  The Dirac delta in frequencies appears due to energy conservation. On the other hand, the product of the spatial delta functions is found by performing the real space loop integrals, which enforce the external points to be coincidental. Once again $g_N$ is a meromorphic function in the $N$ dimensional frequency space with isolated poles, whose detailed structure is unimportant for the proof of the Ward identity. Hence, the above logic of vanishing of \eqref{vanish} also applies here.} 
  
\begin{figure}[h]
			\centering
			\includegraphics[scale=0.175]{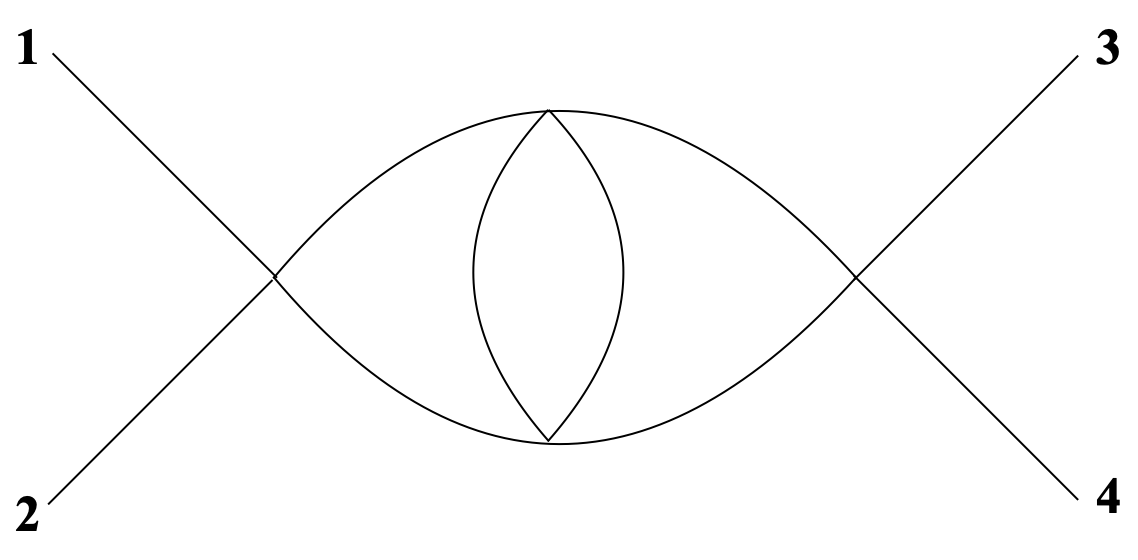}
			\caption{A 3-loop diagram, 4th order in perturbation theory for $\phi^4$ interaction.}
			\label{eye}
	\end{figure}

  One can readily verify this, for example for the 3-loop diagram in Fig. \ref{eye} appearing at 4th order in perturbation theory for $\phi^4$ theory.
          
	\section{Comparison of the beta functions}\label{appendix1}
 
		In this section, we compare the beta functions of the Carrollian and Lorentzian field theories. The results are summarised in Table \ref{table2} and Table \ref{table3} below.
		\begin{table}[h]
			\begin{tabular}{|p{7cm}| |p{7cm}|}
				\hline
				\bf $d=4$ Lorentz/ Euclidean case & \bf $d=4$ Carrollian \\
				\hline
				\bf Action:
				$$ \int d^4x \left[\dfrac{1}{2}\partial_{\mu}\phi \partial^{\mu} \phi+\dfrac{1}{2}{m_0}^2\phi^2+\dfrac{\lambda_{0}}{4!}\phi^4\right] $$ & 
				\bf Action: $$ \int d^4x \Big[\dfrac{1}{2}\partial{_t}\phi\partial{_t}\phi+\dfrac{1}{2}{m_0}^2\phi^2 \
				+\dfrac{\lambda_{0}}{4!}\phi^4\Big]  $$ \\
				
				\hline
				\bf Beta function($\beta_m$):  $$ \dfrac{dm^2}{ds} = -2m^2-\dfrac{\lambda}{16\pi^2}\dfrac{\Lambda^4}{\Lambda^2+m^2}$$ &
				\bf Beta function($\beta_m$): $$ \dfrac{dm^2}{ds} = -2m^2-\dfrac{\lambda}{2\pi}\dfrac{1}{a^3}\dfrac{\Omega}{\Omega^2+m^2}   $$ \\
				\hline
				\bf Beta function($\beta_\lambda$):$$\frac{d\lambda}{ds}=\dfrac{3\lambda^2}{16\pi^2}\dfrac{\Lambda^4}{(\Lambda^2+m^2)^2}$$ &  \bf Beta function($\beta_\lambda$):
				$$\frac{d\lambda}{ds}={-3\lambda}+\dfrac{3\lambda^2}{2\pi}\dfrac{1}{a^3}\dfrac{\Omega}{(\Omega^2+m^2)^2}$$ \\
				\hline
			\end{tabular}
			\caption{Comparison of Beta function in $d=4$ dimensions.} \label{table2}
		\end{table}

  \begin{table}[ht]
			\begin{center}
				\begin{tabular}{|p{8cm}| |p{8cm}|}
					\hline
					\bf $d=3$ Lorentz/ Euclidean case & \bf $d=3$ Carrollian\\
					\hline
					\bf Action: $$ \int d^3x \Big[\dfrac{1}{2} \partial_{\mu}\phi \partial_{\mu}\phi+\dfrac{1}{2}{m_0}^2\phi^2+\dfrac{{\lambda_0}}{4!}\phi^4+\dfrac{\tilde{\lambda}_{0}}{6!}\phi^6\Big] $$ & \bf Action: $$  \int d^3x \Big[\dfrac{1}{2} \partial{_t}\phi \partial{_t}\phi+\dfrac{1}{2}{m_0}^2\phi^2+\dfrac{{\lambda_0}}{4!}\phi^4+\dfrac{\tilde{\lambda}_{0}}{6!}\phi^6\Big] $$ \\
					\hline
					\bf Beta function ($\beta_{m}$):
					$$-2m^2 -\dfrac{{\lambda}}{4\pi^2}\dfrac{\Lambda^3}{(\Lambda^2+m^2)}-\dfrac{ \tilde{\lambda}}{32\pi^2}\dfrac{\Lambda^6}{(\Lambda^2+m^2)^2}$$ & \bf Beta function ($\beta_{m}$):
					$$-2m^2-\dfrac{{\lambda}}{2\pi}\dfrac{1}{a^2} \dfrac{\Omega}{\Omega^2+m^2}- \dfrac{ \tilde{\lambda}}{16\pi^2} \dfrac{1}{a^4} (2s)\dfrac{\Omega^2}{(\Omega^2+m^2)^2}. $$\\
					\hline 
					\bf Beta function ($\beta_{\lambda}$):
					$$\dfrac{d {\lambda}}{ds}={ {-\lambda}}+\dfrac{{3\lambda^2}}{4\pi}\dfrac{\Lambda^3}{(\Lambda^2+m^2)^2}$$ & \bf Beta function ($\beta_{\lambda}$): $$ \dfrac{d {\lambda}}{ds} = {{-3\lambda}}+\dfrac{{{3\lambda}}^2 }{(2\pi)}\dfrac{1}{a^2} \dfrac{\Omega}{(\Omega^2+m^2)^2}$$  \\
					\hline
					\bf Beta function:$$\dfrac{d\tilde{\lambda}}{ds} =\dfrac{\tilde{15\lambda}^2}{32\pi^2}\dfrac{\Lambda^6}{(\Lambda^2+m^2)^3} $$ & \bf Beta function:$$\dfrac{d\tilde{\lambda}}{ds} = {\tilde{-4\lambda}}+ \dfrac{\tilde{15\lambda}^2}{8\pi^2} \dfrac{1}{a^4} (s)\dfrac{\Omega^2}{(\Omega^2+m^2)^3} $$ \\
					\hline 
				\end{tabular}
				\caption{Comparison of Beta function in $d=3$ dimensions.} \label{table3}
			\end{center}
		\end{table} 
  
        \newpage
	\newpage	
		\bibliographystyle{apsrev4-2}
		\bibliography{reference.bib}
\end{document}